\documentclass[apjl]{emulateapj}

\usepackage{natbib}
\usepackage{amsmath, amssymb}
\usepackage{times}
\usepackage{apjfonts, graphicx, graphics}
\usepackage[breaklinks,colorlinks,urlcolor=blue,citecolor=blue]{hyperref}
\usepackage[all]{hypcap}
\usepackage{aas_macros}
\usepackage{subfigure}
\newcommand{\tctff}{{t$_{\rm c}$/t$_{\rm ff}$}}

\shorttitle{A Universal Entropy Profile}
\shortauthors{Iu.V. Babyk et al.}

\begin{document}

% TITLE PAGE
\title{A Universal Entropy Profile for the Hot Atmospheres of Galaxies and Clusters within $R_{2500}$}
\author{Iu.~V. Babyk$^{1,6,\ast}$}
\author{B.~R. McNamara$^{1,2}$}
\author{P.~E.~J. Nulsen$^{3,4}$}
\author{H.~R. Russell$^{5}$}
\author{A.~N. Vantyghem$^{1}$}
\author{M.~T. Hogan$^{1,2}$}
\author{F.~A. Pulido$^{1}$}

\affil{
    $^{1}$Department of Physics and Astronomy, University of Waterloo, Waterloo, ON, N2L 3G1, Canada \\
    $^{2}$Perimeter Institute for Theoretical Physics, Waterloo, ON, N2L 2Y5, Canada \\
    $^{3}$Harvard-Smithsonian Center for Astrophysics, 60 Garden Street, Cambridge, MA 02138, USA \\
    $^{4}$ICRAR, University of Western Australia, 35 Stirling Hwy, Crawley, WA 6009, Australia \\
    $^{5}$Institute of Astronomy, Madingley Road, Cambridge CB3 0HA, UK \\
    $^{6}$Main Astronomical Observatory of the National Academy of Science of Ukraine, 27 Zabolotnogo Street, Kyiv, 03143, Ukraine\\
}

\begin{abstract}
We present atmospheric gas entropy profiles for 40 early type galaxies and 110 clusters spanning several decades of halo mass, atmospheric gas mass, radio jet power, and galaxy type. We show that within $\sim 0.1R_{2500}$ the entropy profiles of low-mass systems, including ellipticals, brightest cluster galaxies, and spiral galaxies, scale approximately as $K\propto R^{2/3}$.  Beyond $\sim 0.1R_{2500}$ entropy profiles are slightly shallower than the $K \propto R^{1.1}$ profile expected from gravitational collapse alone, indicating that heating by AGN feedback extends well beyond the central galaxy.  We show that the $K\propto R^{2/3}$ entropy profile shape indicates that thermally unstable cooling is balanced by heating where the inner cooling and free-fall timescales approach a constant ratio.   Hot atmospheres of elliptical galaxies have a higher rate of heating per gas particle compared to central cluster galaxies.  This excess heating may explain why some central cluster galaxies are forming stars while most early-type galaxies have experienced no significant star formation for billions of years.  We show that the entropy profiles of six lenticular and spiral galaxies follow the $R^{2/3}$ form.  The continuity between central galaxies in clusters, giant ellipticals, and spirals suggests perhaps that processes heating the atmospheres of elliptical and brightest cluster galaxies are also active in spiral galaxies.   
%The origin of the $R^{2/3}$ form is apparently linked to the central galaxy.

\end{abstract}


\altaffiltext{*}{
    \href{mailto:babikyura@gmail.com}{babikyura@gmail.com}
}
\section{Introduction}

Elliptical galaxies, groups, and rich clusters of galaxies are permeated by hot, tenuous atmospheres of plasma that shine in X-rays \citep{Fabian:94}. Hot atmospheres are composed primarily of ionized hydrogen and helium enriched in heavy elements to levels of approximately one third of the Solar value. At their centers, the radiative cooling times fall below $\sim 10^9$ yr, which is much shorter than their ages.  Unless heated, hot atmospheres are expected to cool rapidly into cold molecular clouds fueling star formation and powering active nuclei through accretion onto massive nuclear black holes.  Maintained in hydrostatic equilibrium with thermal pressure balancing gravity, hot atmospheres capture the energy released by radio jets and the heat and metals ejected by supernova explosions. Hot atmospheres are often contain large scale cavities or bubbles inflated by radio jets launched from massive nuclear black holes \citep{McNamara:07, McNamara:12, Fabian:12}. The energy released as bubbles rise through the galaxy, heats the hot atmosphere and prevents catastrophic cooling. 
%The nature of the radiative cooling and heating, through the impact of AGN feedback, as one of the most plausible non-gravitational process occurred at cores of wide-range-mass systems, are not well known. 

%The understanding AGN feedback is significant due to the implications regarding scaling relations and the formation of cosmic structures used in cosmological probes \citep{Cavagnolo:09}. In fact, the studies of X-ray scaling relations have been provided strong evidence of the presence of non-gravitational effects, such as cooling and heating, in cluster cores (their BCGs), isolated, massive early-type galaxies (elliptical and lenticular) and group cores (BGGs) as well \citep{Navarro:95, Navarro:97, Navarro:10, Kravtsov:06, Nagai:07, Boroson:10, Kim:13, Kim:15, Babyk:17scal}. Recent numerical simulations of scaling relations with AGN feedback models have  been provided an additional evidence of the presence of non-gravitational processes within cluster-group-galaxy cores \citep{Borgani:02, Bower:06, Bower:08, Croton:06, Saro:06, Gaspari:12, Gaspari:13}. All these processes can provide an additional heating at the cores of hot atmosphere structures. Thus, the origin this heating in the hot atmospheres of wide-mass-scale systems through the influence of AGN feedback is a focus of this paper.

The history of cooling and heating by AGN and supernova explosions  is encoded in the atmospheric gas entropy \citep{Voit:02, Voit:03, Voit05}. 
%Individual measurements of temperature and density of hot gas do not fully describe a thermal history of hot atmospheres. The temperature of hot atmospheres reflects the potential depth, whilst the density reflects the ability to compress the hot atmosphere gas \citep{Voit05, Cavagnolo:09}. However, the density of hot atmosphere gas within constant pressure can be defined by its specific entropy, $K = T_X n_e^{-2/3}$. The entropy, $K$, describes the thermal history of hot gas since only losses and gains of the heat energy can vary $K$ \citep{Cavagnolo:09}. 
Radial entropy profiles are expected to scale as  $K \propto r^{1.1}$ where the assembly of hot atmospheres is influenced by gravity alone \citep{Kaiser:86, Kaiser:91, Tozzi:01, Voit:02, Voit:03, Voit:05, Voit05, Reiss:15, Babyk:16}.  Departures from this scaling are sensitive to cooling and non-gravitational heating.  The radial entropy profile slopes of cluster atmospheres are shallower than  $K \propto r^{1.1}$  at  $\leq 0.1r_{vir}$ \citep{Ponman:03, Donahue:05, Donahue:06, Pratt:06, Cavagnolo:09, Walker:12, Babyk:14, Panagoulia:14, Hogan:17a}. 
%and that the core entropy shows much bigger uncertainties than the entropies at $\geq 0.1r_{vir}$ radii \citep{Voit:05, Babyk:16}. 
The entropy profiles of groups and massive elliptical galaxies also have shallower central slopes \citep{Werner:12, Werner:13, Voit:15}, indicating complex thermodynamic histories.

\begin{figure}    
\centering
\includegraphics[width=0.49\textwidth]{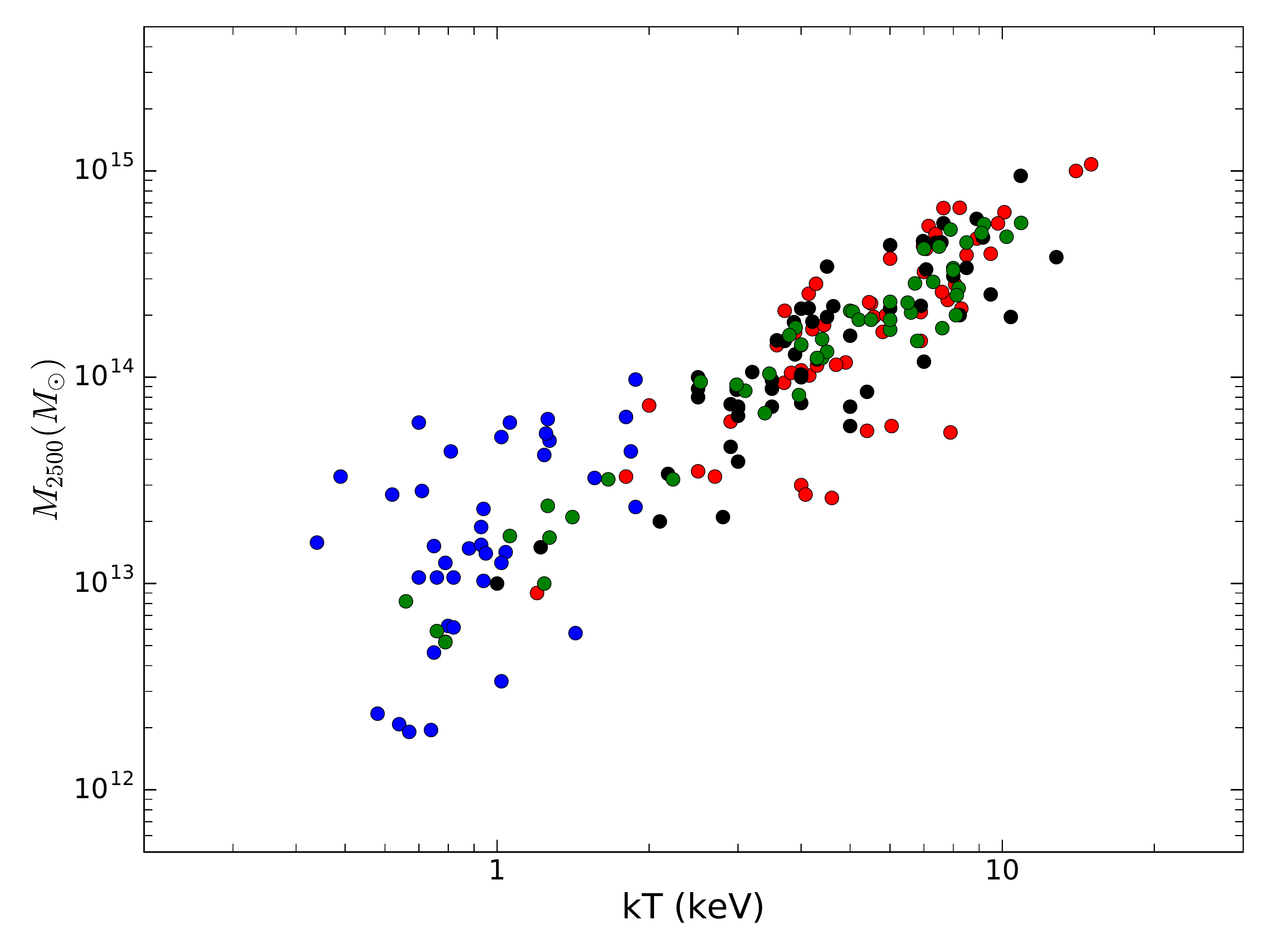}
\caption{\footnotesize The $M_{2500}-T$ relation for the sample of 150 objects spans a wide range of mass and temperature. Data were taken from \citet{Hogan:17a} (red points), \citet{Pulido:17} (black points), \citet{Babyk:17prof} (blue points). In addition, we added measurements of \citet{Main:17} (green points) which were obtained for the same objects used in this paper.}
\label{fig_mt}
\end{figure}

Previous studies revealed a flattening of inner entropy profiles in clusters \citep{Donahue:05, Donahue:06, Cavagnolo:09, Voit:16}.  By carefully tending to resolution biases, \citet{Panagoulia:14} were the first to show that the broken power law fit where the inner entropy profile follows the form $K \propto r^{0.67}$.  Their results were confirmed by 
\citet{Hogan:17a}.  In this paper we present atmospheric entropy profiles for 150 systems observed with the $Chandra$ X-ray Observatory.    Forty are early-type galaxies (elliptical and lenticular galaxies), spiral galaxies, and faint groups. These low-mass systems were combined with 110 central cluster galaxies from \citet{Hogan:17a} and \citet{Pulido:17}.  We examine the thermodynamic states of hot atmospheres permeating halos with masses between 10$^{12}$ to 10$^{15}$ solar masses and atmospheric gas temperatures spaning $0.4-15$ keV (Figure~\ref{fig_mt}). The radio jet powers of the systems shown in Figure~\ref{fig_mt} span $\sim 10^{40}-10^{46}~\rm erg~s^{-1}$. The full X-ray analysis as well as the measurements of thermodynamic properties of low-mass systems, such as temperature, density, cooling time, atmospheric gas mass, and total mass profiles (gas+stars+dark matter) over the radial range $\sim$ 0.1--50.0 kpc, are presented in \citet{Babyk:17prof}.   We concentrate here on the study of entropy profiles obtained in \citet{Babyk:17prof}.

A $\Lambda$CDM cosmology with $\Omega_{\rm m}=0.3$, $\Omega_{\Lambda}=0.7$, and $H_0=70~{\rm km}~{\rm s}^{-1}~{\rm Mpc}^{-1}$ is adopted.  Errors are quoted at the 1$\sigma$ confidence level, unless otherwise specified.

\section{Observations and data reduction} 

The sample is drawn from our parent sample \citep[see][]{Babyk:17scal}, where 94 early-type galaxies (ETGs) and other low-mass systems observed by the {\it Chandra} X-ray Observatory were analyzed to investigate X-ray scaling relations. Combining multiple observations where necessary, we selected 40 targets with exposure times exceeding $10$~ks and having sufficient counts to build spatially resolved thermodynamic profiles. This selection were performed in \citet{Babyk:17prof} where we construct radial profiles of main thermodynamic properties, including temperature, density, cooling time, and mass. Characteristics of the selected targets are presented in Table~\ref{tab1}. Complimentary information was obtained from NED\footnote{https://ned.ipac.caltech.edu/}, SIMBAD\footnote{http://simbad.u-strasbg.fr/}, and HyperLEDA\footnote{Lyon-Meudon Extragalactic Database} databases. Our sample includes 11 brightest cluster galaxies and several spiral galaxies. Angular diameter and luminosity distances were calculated using redshifts or surface brightness fluctuations \citep{Mei:07}, as appropriate. 

\begin{table*}
\fontsize{8}{8}
\selectfont
\caption{Selected low-mass systems.}\label{tab1}
\centering
\begin{tabular}{lcccccccccccc}
\hline
 && \\
Name      & $\alpha$ & $\delta$ & ObsIDs & Exposure & Type & BCG  &  $z$     &  $D_A$ &  $D_L$  &  $N_{H}$ & $\sigma_c$ \\
          &  (J2000)   & (J2000)    &       &    ks     &      &         &          &   Mpc  &  Mpc    &  10$^{20}$ cm$^{2}$ & km/s \\
          & (2) & (3) & (4) & (5) & (6) & (7) &(8) & (9) & (10) & (11) & (12) \\
&& \\
\hline
&&\\
IC1262  & 69.5188 & 32.0738 & 6949, 7321, 7322 & 36.02, 34.98, 35.17 & E & $\surd$ & 0.032649 & 133.0 & 141.8 & 2.47 & 232$\pm$10 \\
IC1459  & 4.6590 & -64.1096   & 2196 & 45.14 & E3 & & 0.006011 & 25.503 & 25.8 & 1.19 & 294$\pm$6 \\
IC4296  & 313.5384 & 27.9729 & 2021, 3394 & 19.27, 20.78 & E &  & 0.012465 & 52.358 & 53.7 & 4.11 & 327$\pm$5 \\
NGC315  & 124.5631 & -32.4991 & 4156 & 39.49 & E &         & 0.016485 & 68.816 & 71.1 & 5.87 & 293$\pm$2 \\
NGC499  & 130.4977 & -28.9448 & 10536 & 18.33 & E & & 0.014673 & 61.423 & 63.2 & 5.26 & 253$\pm$7 \\
        & && 10865 & 5.12 & \\
        & && 10866 & 8.01 & \\
        & && 10867 & 7.02 & \\
NGC507  & 130.6430 & -29.1326 & 317 & 40.30 & E & $\surd$ & 0.016458 & 68.706 & 71.0 & 5.32 & 292$\pm$6 \\
NGC533  & 140.1457 & -59.9683 & 2880 & 28.40 & E3 & & 0.018509 & 77.025 & 79.9 & 3.12 & 271$\pm$6 \\
NGC708  & 136.5695 & -25.0903 &  2215, 7921 & 28.75, 108.63   & E    & $\surd$ & 0.016195 & 67.635 & 69.8  &  5.37 & 222$\pm$8\\
NGC720  & 173.0194 & -70.3572 & 7372 & 49.13 & E5 &    & 0.005821 & 24.704 & 25.0 & 1.55 & 236$\pm$6\\
        & && 7062 & 22.12 & \\
        & && 8448 & 8.06 & \\
        & && 8449 & 18.91 & \\
NGC741  & 150.9342 & -53.6764 & 2223 & 28.14 & E0 &         & 0.018549 & 77.186 & 80.1 & 4.47 & 286$\pm$9 \\
NGC1316 & 240.1627 & -56.6898 & 2022  & 21.21  & E &         & 0.005871 & 24.914 & 25.2  & 1.92 & 224$\pm$3\\
NGC1332 & 212.1830 & -54.3661 & 2915, 4372  & 4.10, 16.38  & S0  &         & 0.005084 & 21.601 & 21.8  & 2.29 & 313$\pm$11 \\
NGC1399 & 236.7164 & -53.6356 & 9530  & 56.98    & E1   & $\surd$ & 0.004753 & 20.205 & 20.4  &  1.31 & 334$\pm$5\\
NGC1404 & 236.9552 & -53.5548 & 16233 & 91.94  & E1  & & 0.006494 & 27.531 & 27.9  & 1.35 & 228$\pm$4 \\
        & && 16231 & 56.09 &\\
        & && 16232 & 64.03 &\\
        & && 16234 & 84.64 &\\
NGC1407 & 209.6362 & -50.3838 & 14033 & 50.26    & E0   &         & 0.005934 & 25.179 & 25.5  &  5.41 & 265$\pm$5 \\
NGC1550 & 190.9760 & -31.8488 & 5800, 5801 & 44.55, 44.45 & E2   & $\surd$ & 0.012389 & 52.045 & 53.3 & 11.2  & 300$\pm$5 \\
NGC3091 & 256.7559 & 27.5029 & 3215 & 27.34 & E3 & $\surd$ & 0.013222 & 55.473 & 56.9 & 4.75 & 310$\pm$7 \\
NGC3923 & 287.2759 & 32.2224  & 9507  & 80.90  & E4  &         & 0.005801 & 24.620 & 24.9 & 6.29 & 247$\pm$6 \\
NGC4073 & 276.9081 & 62.3697 & 3234 & 25.76 & E  & $\surd$ & 0.019584 & 81.364 & 84.6 & 1.90 & 267$\pm$6 \\
NGC4104 & 204.3284 & 80.0306 &  6939 & 34.86    & S0  & $\surd$ & 0.028196 & 115.60  & 122.2 &  1.68 & 291$\pm$6 \\
NGC4125 & 130.1897 & 51.3391  & 2071   & 52.97 & E6  &         & 0.004523 & 19.234 & 19.4 & 1.86 & 238$\pm$7 \\
NGC4261 & 281.8049 & 67.3726 &  9569 & 102.24   & E2 &         & 0.007378 & 31.236 & 31.7  &  1.56 & 296$\pm$4 \\
NGC4325 & 279.5840 & 72.1969 & 3232 & 28.30 & E4 & $\surd$ & 0.025714 & 105.80 & 111.3 & 2.18 & 299$\pm$78 \\
NGC4374 & 278.2045 & 74.4784  & 5908, 6131 & 44.04, 35.81  & E1 &           & 0.003392 & 16.422 & 18.1 & 2.58 & 275$\pm$2 \\
NGC4382 & 267.7120 & 79.2372  & 2016 & 29.33  & S0 &          & 0.002432 & 16.265 & 17.7 & 2.51 & 175$\pm$4  \\
NGC4472 & 286.9222 & 70.1961 & 11274 & 39.67    & E2   &         & 0.003272 & 15.621 & 17.1  &  1.65 & 281$\pm$3\\
NGC4552 & 287.9326 & 74.9668  & 13985 & 49.41 & E  &           & 0.001134 & 15.523 & 16.1 & 2.56 & 250$\pm$3 \\
        & && 14358 & 49.41 & \\
        & && 14359 & 47.11 & \\
NGC4636 & 297.7485 & 65.4729 & 3926, 4415  & 67.26, 66.17    & E0   &         & 0.003129 & 13.335 & 13.4  &  1.83 & 200$\pm$3  \\
NGC4649 & 295.8736 & 74.3178 & 8182, 8507  & 45.87, 15.73    & E2   &         & 0.003703 & 15.767 & 15.9  &  2.13 & 329$\pm$5 \\
NGC4696 & 302.4036 & 21.5580 & 1560  & 21.20 & E1 & $\surd$ & 0.009867 & 41.613 & 42.4 & 8.07 & 244$\pm$6 \\
NGC4782 & 304.1379 & 50.2958 & 3220 & 49.33 & E0 &          & 0.015437 & 64.545 & 66.6 & 3.56 & 308$\pm$11 \\
NGC5044 & 311.2340 & 46.0996 & 17195 & 77.01    & E0   & $\surd$ & 0.00928  & 39.173 & 39.9  &  5.03 & 226$\pm$9  \\
        &          &          & 17196 & 85.80    &      &         \\
        &          &          & 17653 & 32.46    &      &         \\
        &          &          & 17654 & 24.01    & \\
        &          &          & 17666 & 82.79    & \\
NGC5353 & 82.6107 & 71.6336 & 14903 & 37.20 & S0 &            & 0.007755 & 32.813 & 33.3 & 0.98 & 284$\pm$5 \\
NGC5813 & 359.1820 & 49.8484 & 12952 & 140.00 & E1 &        & 0.006525 & 27.662 & 28.0 & 4.23 & 235$\pm$3 \\
        &          &          & 12951 & 71.95    &  \\
        &          &          & 12953 & 31.76    & \\
        &          &          & 13246 & 45.02    & \\
        &          &          & 13247 & 34.08    & \\
        &          &          & 13255 & 43.34    & \\
NGC5846 & 0.3389   & 48.9043 &  7923 & 85.25    & E    &         & 0.00491  & 20.867 & 21.1  &  4.24 & 237$\pm$4 \\
NGC6338 & 85.8062 & 35.3991 & 4194 & 44.52 & E5 &           & 0.027303 & 112.10 & 118.3 & 2.55 & 348$\pm$40 \\
NGC6482 & 48.0905 & 22.9122 & 3218 & 10.03 & E &            & 0.013129 & 55.091 & 56.5 & 8.04 & 317$\pm$10 \\
NGC6861 & 350.8772 & -32.2109 & 11752 & 88.89 & SA0 &        & 0.009437 & 39.826 & 40.6 & 4.94 & 407$\pm$20 \\
NGC7618 & 105.5754 & -16.9091 & 16014 & 121.00   & E &         & 0.017309 & 72.164 & 74.7  &  11.9 & 293$\pm$30 \\
UGC408  & 116.977 & -59.40 & 11389 & 93.80 & SAB & & 0.014723 & 61.628 & 63.5 & 2.80 & 198$\pm$5 \\
\hline
\end{tabular}
\end{table*}

The $Chandra$ X-ray data have been reduced following \citet{Hogan:17a} and \citet{Pulido:17}, summarized here. The data were reprocessed and bad pixel files were created using the {\sc ciao v.4.8} software package with the newest version of {\sc caldb v.4.7.1}. \texttt{chandra\_repro} was used to extract the cleaned, level-2 event files. Background flares were removed and a correction was applied for time-dependent gains. Point sources were identified using \texttt{wavdetect} and removed. Spectra were then extracted from concentric circular annuli, avoiding bubbles and other asymmetric features, and were deprojected using the {\sc dsdeproj} routine \citep{Sanders:07, Russell:08}. 

We use a multi-component spectral model of the form {\sc phabs*(apec+po+mekal+po)} and {\sc xspec} version 12.9.1 \citep{Arnaud:96} to fit the deprojected spectra. The spectral model included  thermal X-ray emission from the galaxy and its environment ({\sc apec} component in our spectral model), unresolved low-mass X-ray binaries (LMXB) (first {\sc po} component with slope fixed to 1.6), and other stellar sources (set of {\sc mekal+po} components with temperature, metallicity, and slope fixed to 0.5 keV, 0.3$Z/Z_{\odot}$, and 1.9, respectively) \citep[see][for more details]{Babyk:17scal, Babyk:17prof}. X-ray emission from LMXBs and other stellar sources were modeled as power laws along with the thermal emission from the galaxy. The flux contributed by these sources was usually negligible compared to the thermal emission. However, we study faint systems. Thus, the magnitude and spectral form of the contributions of stellar sources and unresolved low-mass X-ray binaries must be assumed \citep{Boroson:10, Kim:13, Kim:15}. From these fits we derive temperature and electron density profiles which we then use to build entropy profiles.

Clusters with relatively short central cooling times are better fit in the inner regions by two temperature rather than single-temperature thermal models \citep{Panagoulia:14, Hogan:17a}. \citet{Hogan:17} and \citet{Hogan:17a} showed that two temperature models are only useful in observations with sufficiently high numbers of photons, and usually when measuring temperatures in projection. {We tested the spectra for a second temperature component. Galaxies were refit with a two-temperature model in the form of {\sc phabs*(apec+apec+po+mekal+po)}. No evidence for a second temperature component was found in any system.} As we focus here on deprojected radial profiles of faint systems, we have attempted to fit only single temperature thermal models.

\section{Modelling the entropy profiles} 

\begin{figure*}
\centering
\includegraphics[width=0.47\textwidth]{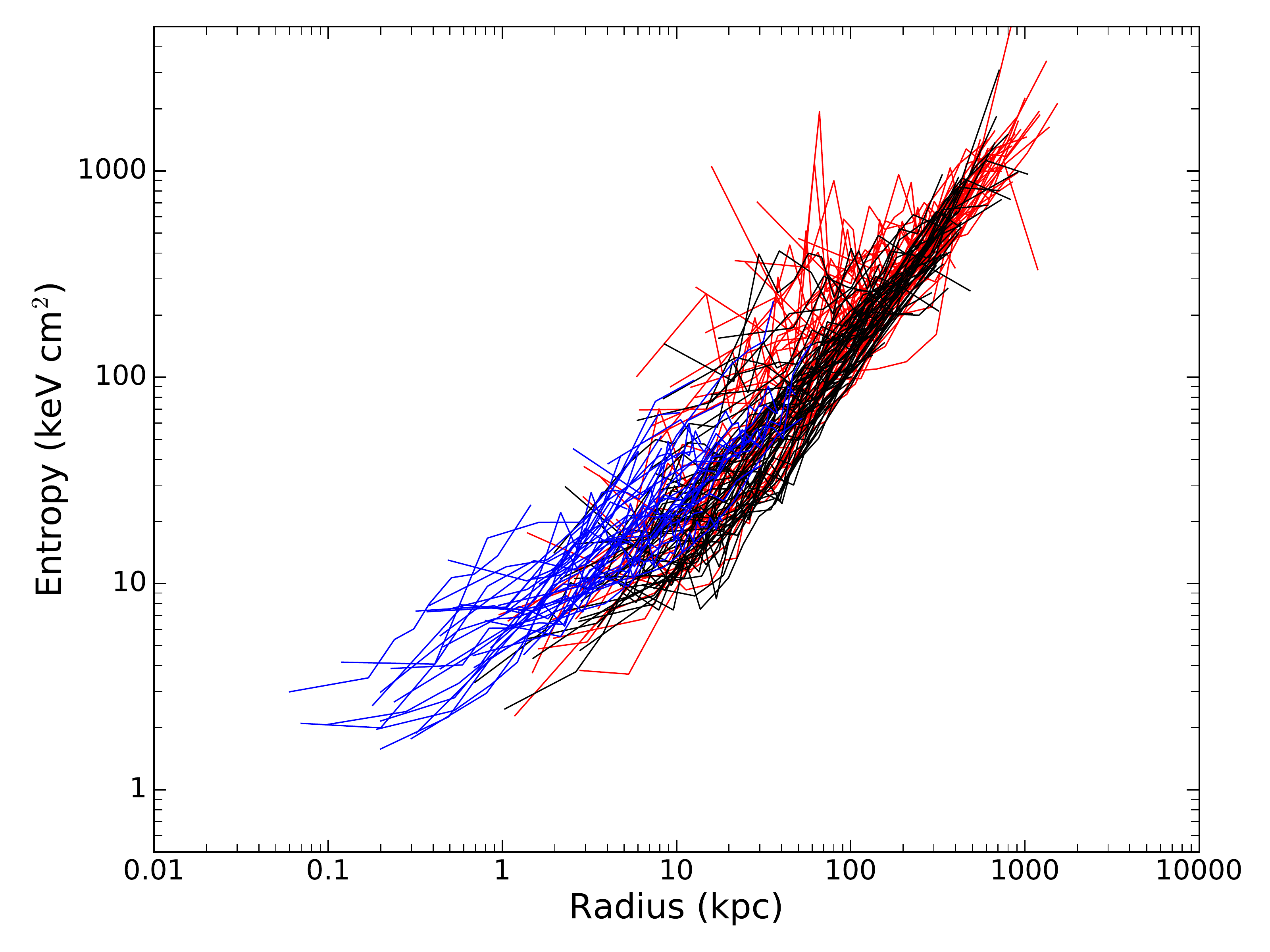}
\includegraphics[width=0.47\textwidth]{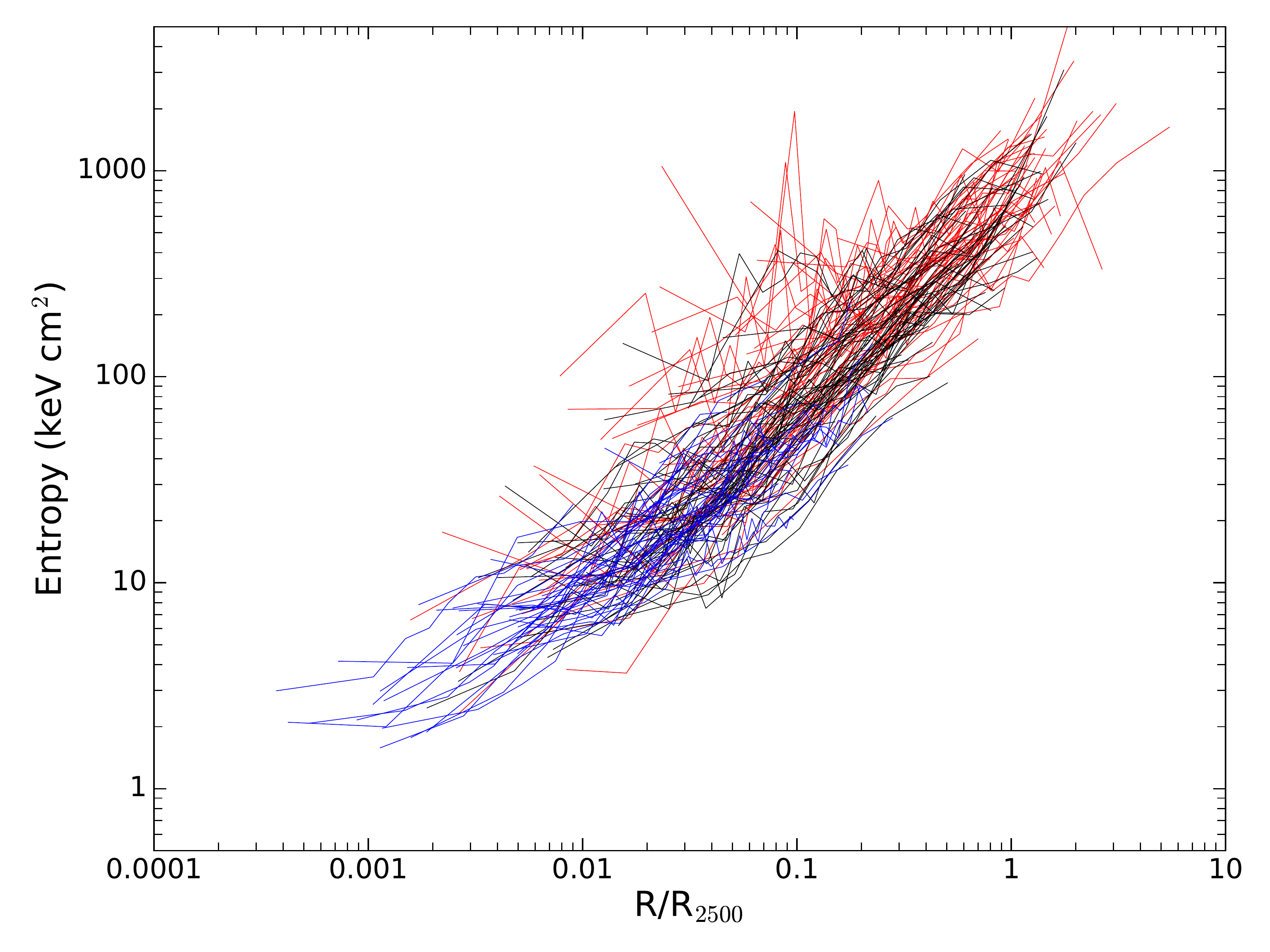}
\caption{Entropy profiles for the sample of clusters, groups and galaxies vs. radius (left) and scaled radius (right).  The red profiles are taken from \citet{Hogan:17a}, the black profiles are from \citet{Pulido:17}, and the blue profiles are from \citet{Babyk:17prof}. Error bars have been omitted for clarity.}
\label{fig_2}
\end{figure*}

Radial gas entropy profiles were calculated as $K(r)=kT(r)/n_e(r)^{2/3}$, where $kT$ and $n_e$ are the gas temperature and electron density determined by deprojecting a single thermal emission model.  The gas entropy profiles for the low-mass systems were compared to those of central cluster galaxies. Entropy profiles of 150 halos are shown in Figure~\ref{fig_2}. Early-type galaxies/faint groups are shown in blue while clusters are shown in black \citep{Pulido:17} and red \citep{Hogan:17a}. This figure shows a mean entropy profile over four decades in radius, from the inner 100 parsecs in nearby early-type galaxies to beyond a megaparsec in some galaxy clusters.  The right hand panel shows the entropy profile with the radius in units of $R_{2500}$ (within $R_{2500}$ the mean total density is 2500 times the cosmological critical density). 

Entropy profiles, in both physical and scaled radial units, extend continuously from galaxy to cluster scales. The low-mass systems and clusters were first modelled separately using a broken power of the form
\begin{equation}
 f(x) = \begin{cases}
  A\cdot(x/x_0)^{\Gamma_1} & \quad \text{if} \quad x \leq x_0 \\
  A\cdot(x/x_0)^{\Gamma_2} & \quad \text{if} \quad x > x_0.
 \end{cases}
 \label{eq1}
\end{equation}

The entropy profiles in individual galaxies, faint groups, and the centers of clusters are shallower than those at larger radii in clusters. Comparing entropy profiles in physical radii, the normalizations of low-mass systems were found to be 1.5 times larger than cluster normalizations. For the subsequent analysis we have divided the entropies for low-mass systems by 1.5 with respect to cluster profiles when plotted against physical radius, as in Figure~\ref{fig_2}. This shifts the break radius of the broken power law fitting slightly, but does not affect slopes. The normalizations of the entropy profiles with scaled radii were not changed.  
%The factor 1.5 scaling is significant and its physical implications are given below. {\bf Due to lower mass, the impact of baryon accretion in groups and galaxies is stronger compared to galaxy clusters. Because of this difference between the accretion in low-mass systems and clusters, their data points trends are characterized by different normalization. \citet{Voit03} found this difference as factor of $\sim$2 which is in consistency with our estimates. In addition, the low-mass systems are able to supply the higher energy per gas particle of heating required to assume the observed entropy. The calculations and discussion related to heating are discussed further below. }

Despite the broad range of mass and spatial resolution, Figures~\ref{fig_2} and \ref{fig_1} show remarkably uniform profile shapes, although with noticeable variations. The degree to which the variations are due to measurement error, non-uniform exposure level, or real departures from a universal form is unclear. Therefore tests were performed to evaluate the degree to which these systems can be characterized by a single inner power law slope.   

\subsection{Scatter about the mean entropy profile}

We focus on broken power law fits to the entropy profiles of low-mass systems and cool-core clusters.  By construction these systems have central cooling times and entropies that fall below 1 Gyr and 30 keV cm$^2$, respectively.  The entropy profiles are presented in physical units and scaled units as $K$ vs $R$, $K$ vs $R/R_{2500}$ and 
$K/K_{2500}$ vs $R/R_{2500}$. To evaluate the variance, profiles were constructed using three methods.  

\begin{figure*}
\centering
\includegraphics[width=0.49\textwidth]{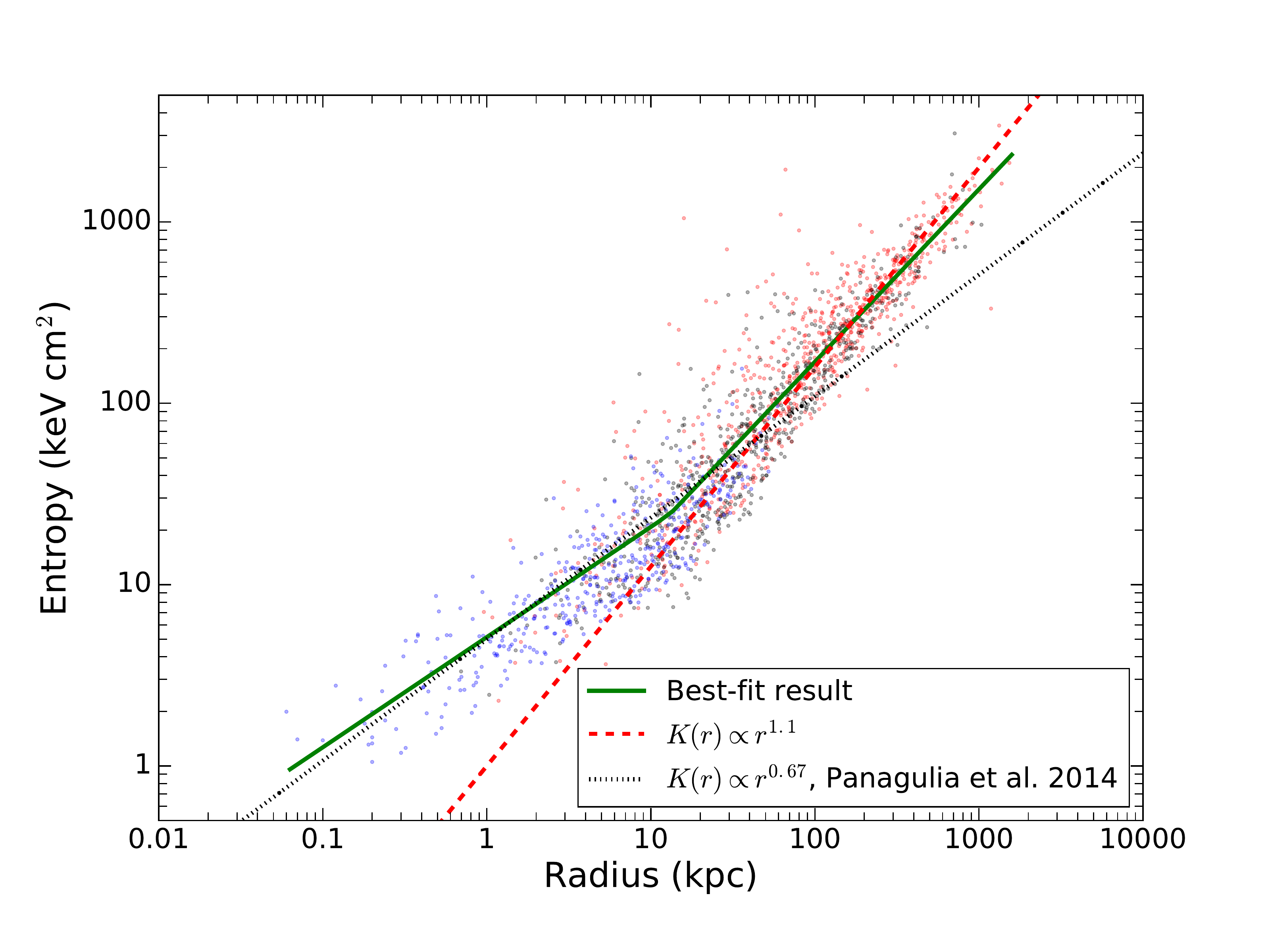}
\includegraphics[width=0.49\textwidth]{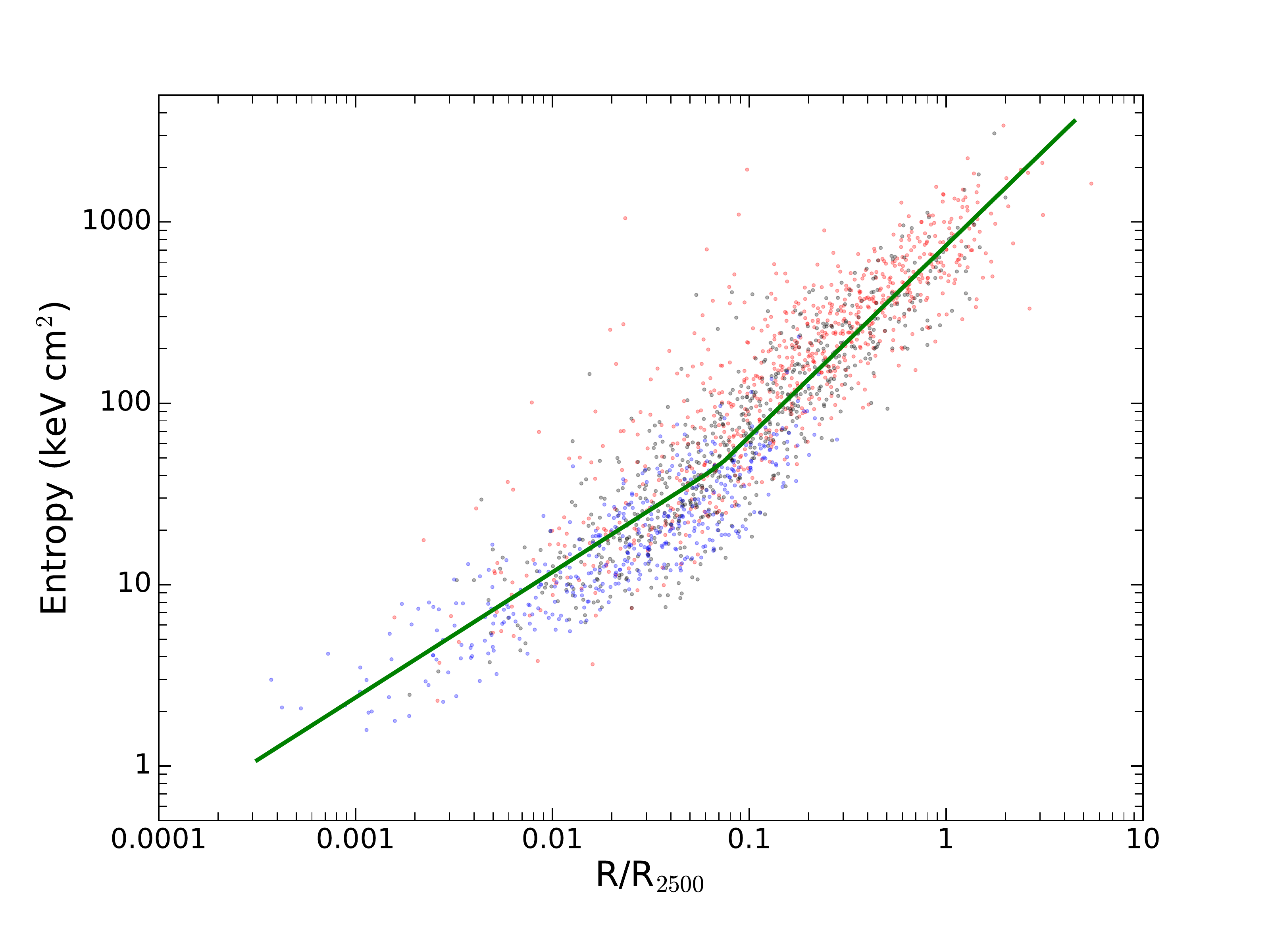}
\caption{\footnotesize The distribution of deprojected radial (left) and radially scaled (right) entropy profiles fitted by the broken power law models discussed in the text. {The blue points correspond to the \citet{Babyk:17prof} entropy profiles while the red and black points to those entropy profiles derived in \citet{Hogan:17a} and \citet{Pulido:17} respectively.} The error bars were deleted for clarity. The red dashed line corresponds to fixed slope $R^{1.1}$ while black dotted line corresponds to a slope of $R^{0.67}$ found by \citet{Panagoulia:14}.}
\label{fig_1}
\end{figure*}

First, a mean profile was constructed by fitting a broken power law to the entropy values vs. radius for the entire sample.  The data along with the best fits in green are shown in Figure~\ref{fig_1}.  We have plotted the inner profile found by \citet{Panagoulia:14} and the outer $K\propto R^{1.1}$ profile for reference.  The agreement is good although the outer profile is slightly shallower than $R^{1.1}$.   

The broken power-law form was fit to the entropy profiles with both physical and scaled radial units using a {simple $\chi^2$ method. Our uncertainties were estimated from the log-likelihood profile following the Wilks theorem. They were computed as the parameter ranges where the log-likelihood does not deviate by more than 1 with respect to its best-fit value.}
%{\bf The uncertainties on individual measurements were been factored into this optimization. }
In physical radii the power law slopes were $\Gamma_1=0.62\pm0.12$ and $\Gamma_2=0.95\pm0.17$, and the break radius was $14.3\pm5.2$~kpc. For scaled radii the slopes were $\Gamma_1=0.69\pm0.09$ and $\Gamma_2=1.05\pm0.14$ and the break radius was $(0.07\pm0.02)~R/R_{2500}$. 

{The full results of fitting for unbinned and binned (see below) data, including inner ($\Gamma_1$) and outer ($\Gamma_2$) slopes, breaks as well as verification test, $\chi^2$, and probability, $p$, are given in Table~\ref{tab_fit}. Three different relations, including $K$ vs $R$, $K$ vs $R/R_{2500}$, and $K/K_{2500}$ vs $R/R_{2500}$ are presented. $K$ vs $R$, $K$ vs $R/R_{2500}$ are shown in Figure~\ref{fig_1}}.

Additional power law fits were performed for entropies above and below 50~kpc to compare with \citet{Panagoulia:14}. The entropy profiles are characterized by a power law slopes of 1.00$\pm$0.16 above 50~kpc and 0.69$\pm$0.09 below. These two additional power law models cross at $\sim$ 40 kpc. In Figure~\ref{fig_1} (left) we show the entropy profiles predicted from gravitational collapse models, $K(r) \propto r^{1.1}$, as well as those measured below 50~kpc obtained in \citet{Panagoulia:14}. Interestingly, \citet{Panagoulia:14} and predicted models cross at $\sim$ 42.5 kpc in agreement with our power law results.

Second, a global mean entropy profile was constructed by averaging the entropy values at each radius, both scaled and physical, for all systems.   
\begin{table*}
\centering
\caption{The best-fit result of broken power law fits (Eq.~\ref{eq1}) to the entropy profiles obtained for entire sample (All) and for low-mass systems plus cool-core clusters only (CC+Galaxies).}\label{tab_fit}
 \begin{tabular}{cccccccc}
  \hline
  &\\
  Relation & $\Gamma_1$ & $\Gamma_2$ & Break & $\chi^2/{\rm dof}$ & $p$ & Data \\
  &                  &           &  kpc     & &\\
  \hline
  $K$ vs $R$                    & 0.62$\pm$0.12 & 0.97$\pm$0.17 & 14.3$\pm$5.2  & 1.5 & $>>$0.0001 & All \\
  $K$ vs $R/R_{2500}$           & 0.69$\pm$0.09 & 1.05$\pm$0.14 & 0.07$\pm$0.02 & 1.4 & $>>$0.0001 & All \\
  $K/K_{2500}$ vs $R/R_{2500}$ & 0.67$\pm$0.09 & 1.03$\pm$0.10 & 0.07$\pm$0.03 & 1.5 & $>>$0.0001 & All \\
  $K$ vs $R$                    & 0.65$\pm$0.11 & 1.02$\pm$0.11 & 16.4$\pm$5.5 & 1.3 & $>>$0.0001 & CC+Galaxies \\
  $K$ vs $R/R_{2500}$           & 0.64$\pm$0.12 & 0.99$\pm$0.15 & 0.07$\pm$0.02 & 1.3 & $>>$0.0001 & CC+Galaxies \\
  $K/K_{2500}$ vs $R/R_{2500}$ & 0.68$\pm$0.13 & 1.02$\pm$0.17 & 0.07$\pm$0.02 & 1.2 & $>>$0.0001 & CC+Galaxies \\
  %$K\propto R$                    & 0.64$\pm$0.05 & 0.95$\pm$0.07 & 11.92$\pm$0.07& 1.4 & NCC+Galaxies \\
  %$K\propto R/R_{2500}$           & 0.68$\pm$0.07 & 1.12$\pm$0.11 & 0.07$\pm$0.02 & 1.5 & NCC+Galaxies \\
  %$K/K_{2500} \propto R/R_{2500}$ & 0.69$\pm$0.08 & 1.10$\pm$0.13 & 0.08$\pm$0.02 & 1.5 & NCC+Galaxies \\
  \hline
  &   &  Binned data & &\\
  \hline
  $K\propto R$                    & 0.68$\pm$0.06 & 0.99$\pm$0.11 & 15.4$\pm$3.6  & 1.1 & 0.36 & All \\
  $K\propto R/R_{2500}$           & 0.62$\pm$0.09 & 1.05$\pm$0.12 & 0.07$\pm$0.02 & 1.2 & 0.41 & All \\
  $K/K_{2500} \propto R/R_{2500}$ & 0.71$\pm$0.07 & 1.05$\pm$0.09 & 0.07$\pm$0.02 & 1.2 & 0.40 & All \\
  $K\propto R$                    & 0.65$\pm$0.06 & 1.02$\pm$0.10 & 16.8$\pm$3.7 & 1.2 & 0.41 & CC+Galaxies \\
  $K\propto R/R_{2500}$           & 0.66$\pm$0.08 & 1.06$\pm$0.08 & 0.07$\pm$0.02 & 1.1 & 0.32 & CC+Galaxies \\
  $K/K_{2500} \propto R/R_{2500}$ & 0.68$\pm$0.09 & 1.02$\pm$0.11 & 0.07$\pm$0.02 & 1.3 & 0.45 & CC+Galaxies \\
  %$K\propto R$                    & 0.67$\pm$0.07 & 1.08$\pm$0.09 & 16.39$\pm$0.26 & 1.4 & NCC+Galaxies \\
  %$K\propto R/R_{2500}$           & 0.69$\pm$0.08 & 1.12$\pm$0.11 & 0.08$\pm$0.02  & 1.5 & NCC+Galaxies \\
  %$K/K_{2500} \propto R/R_{2500}$ & 0.70$\pm$0.09 & 1.10$\pm$0.14 & 0.07$\pm$0.02 & 1.4 & NCC+Galaxies \\
  \hline
  \end{tabular}
  %CC \& NCC are cool-core and non-cool-core clusters
\end{table*}

%\subsection{Global Profile Analysis}

\begin{figure}
\centering
\includegraphics[width=0.48\textwidth]{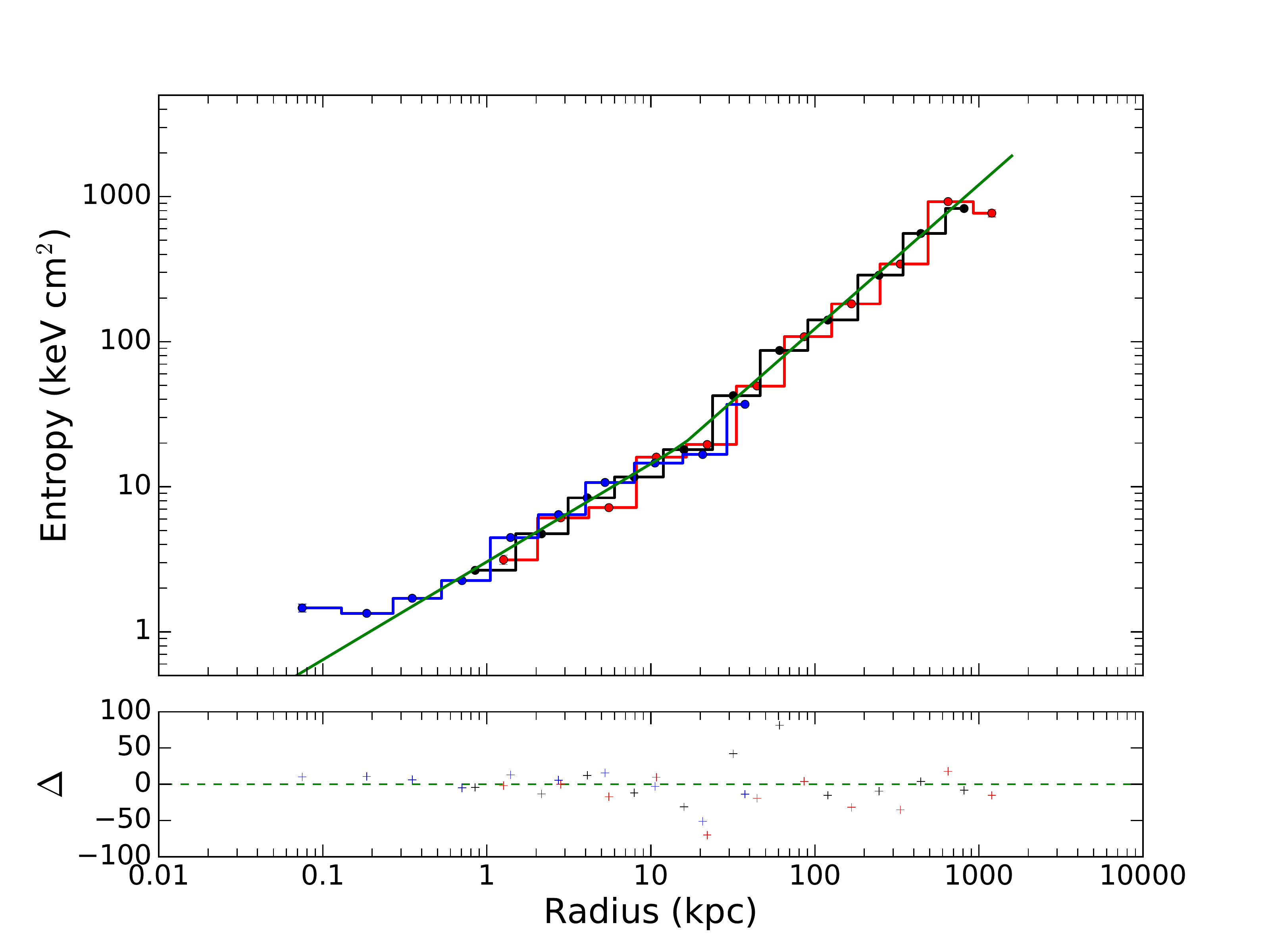}\\
\includegraphics[width=0.48\textwidth]{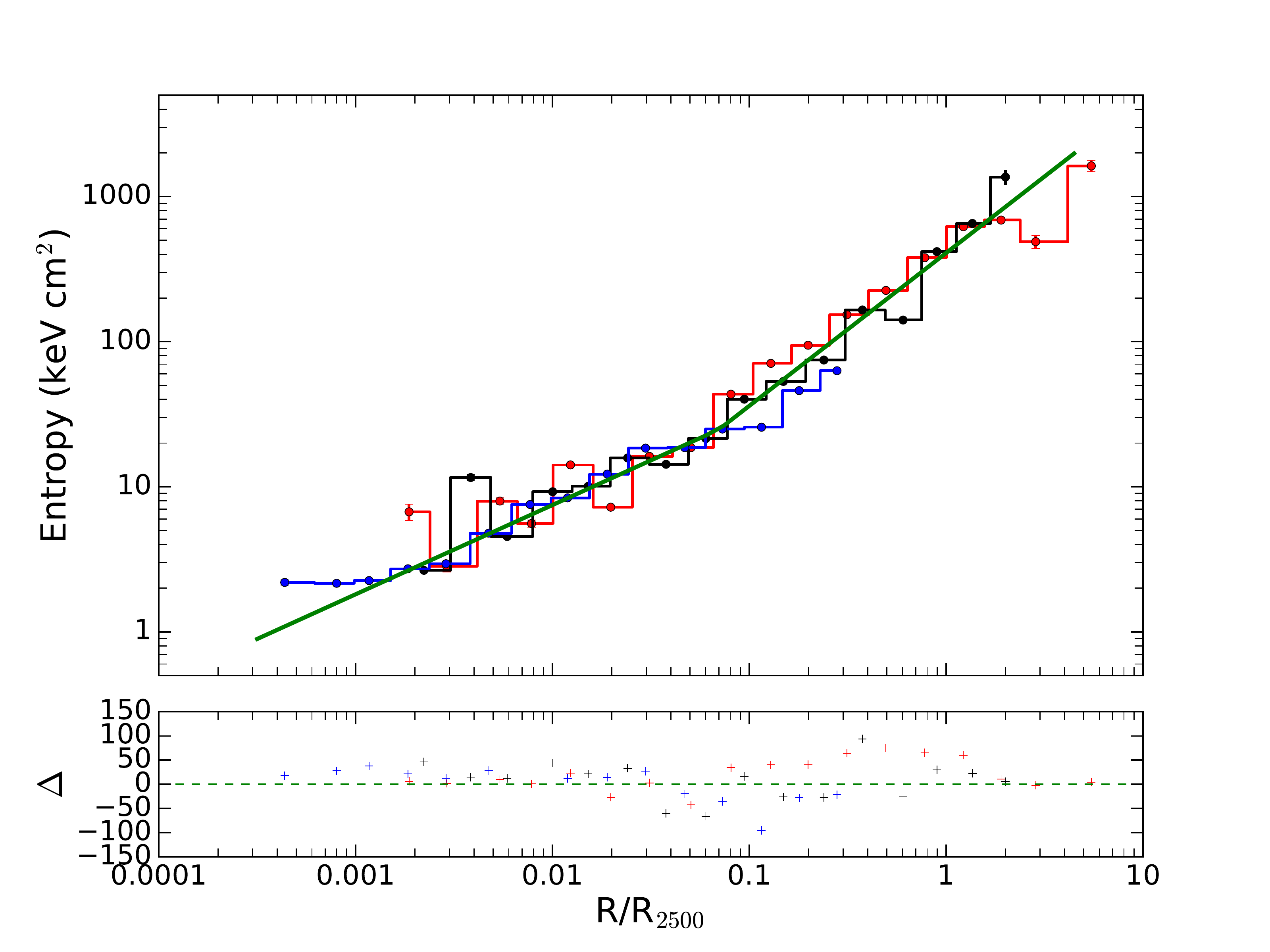}\\
\includegraphics[width=0.48\textwidth]{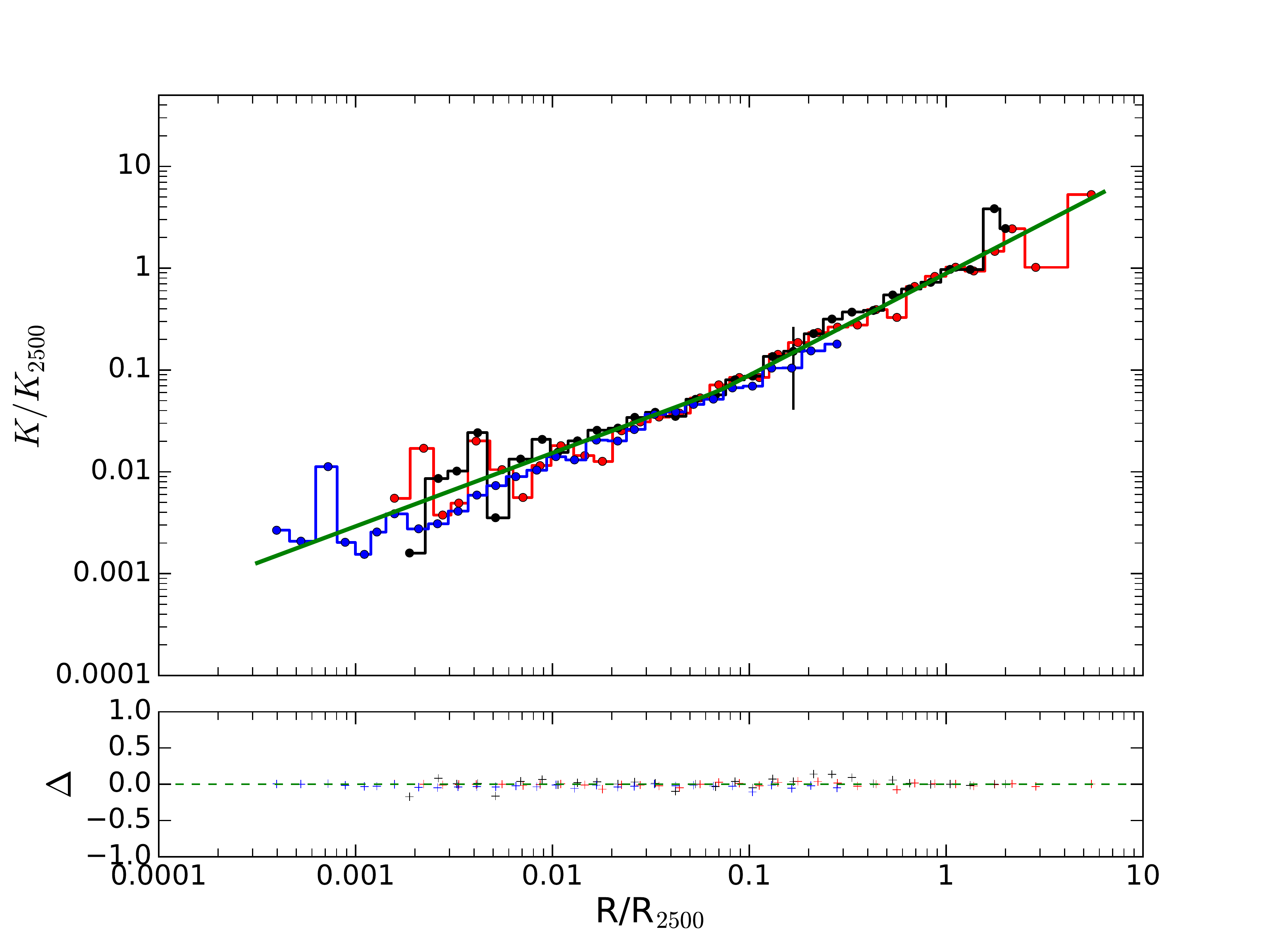}\\
\caption{The broken power law model fitting, including residuals, of the grouped entropy bins (green lines) for three sets of data. Blue points correspond to low-mass systems while red and black to \citet{Hogan:17a} and \citet{Pulido:17} measurements, respectively.}
\label{fig_gr}
\end{figure}

The entropy values at each radius were grouped into bins, $x_i$, and the weighted mean entropy was calculated as $< y > = \frac{\sum_i(y_i\times  \sigma_i^{-2})}{\sum_i \sigma_i^{-2}}$. The uncertainties of the weighted mean entropy were found as $\sigma_{<y>} = \sqrt{\frac{1}{\sum_i \sigma_{i}^{-2}}}$, where the $y_i$ and $\sigma_i$ are entropies and their errors, respectively.  Assuming that $y_i$ are normally distributed and independent, the weighted mean is the maximum likelihood estimator. The interval $x$ is $0.03$ kpc for physical radius and 0.01$R/R_{2500}$ for scaled. The results are presented in Table~\ref{tab_fit} {for three relations and are quoted as ``Binned data''}.  The profiles and residuals for three sets of entropies and best-fit modes are shown in Figure~\ref{fig_gr}.   

Inspection of Table~\ref{tab_fit} shows that both approaches give similar results.  All slopes agree to within their errors.  The mean profile is remarkably tight indicating that the mean slope characterizes the clusters quite well. {The second approach (binning) was performed to understand the effects of scatter in the entropy profiles. Both methods agree.}

The binned data were fitted for all targets simultaneously and for the low-mass systems and cool-core clusters separately.  
The central cooling times of non-cool-core objects exceeds $\sim$ 2 Gyr. The corresponding central entropies exceed $\sim$ 50 keV cm$^2$.  In contrast, the central cooling time and entropy of cool-core clusters lie below 1 Gyr and 30 keV cm$^2$, respectively. Non-cool core systems generally do not host central galaxies.  Table~\ref{tab_fit} shows that outer profiles of cool-core and non-cool core clusters are indistinguishable. Both scale roughly as $K\propto R$.  The inner profiles of the non-cool core clusters are poorly defined and suffer from low count rates and thus poor statistics.  We have therefore not attempted to fit their central regions. Our results for non-cool-core clusters are consistent with previous studies \citep{David:96, Ponman:99, David:01, Ponman:03, Panagoulia:14, Hogan:17a, Pulido:17}. 

\subsection{Entropy distribution of low-mass systems}

Our third approach examines the power-law slopes for each individual low-mass system alone and compares their distribution to the globally-averaged profiles.   This analysis is sensitive to understanding fitting biases and real departures from the mean and perhaps their physical cause.

\begin{figure}
\centering
\includegraphics[width=0.5\textwidth]{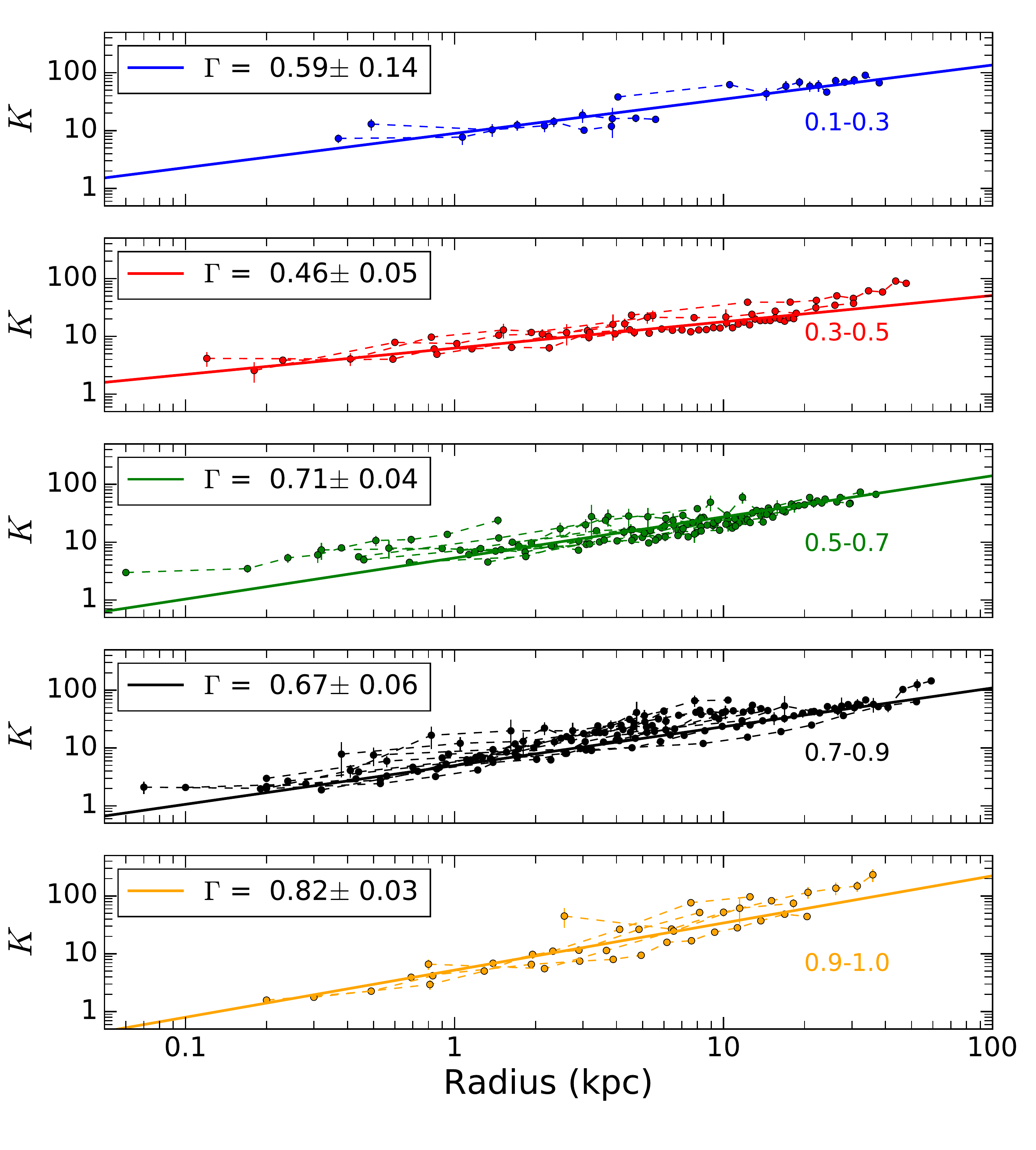}
\includegraphics[width=0.5\textwidth]{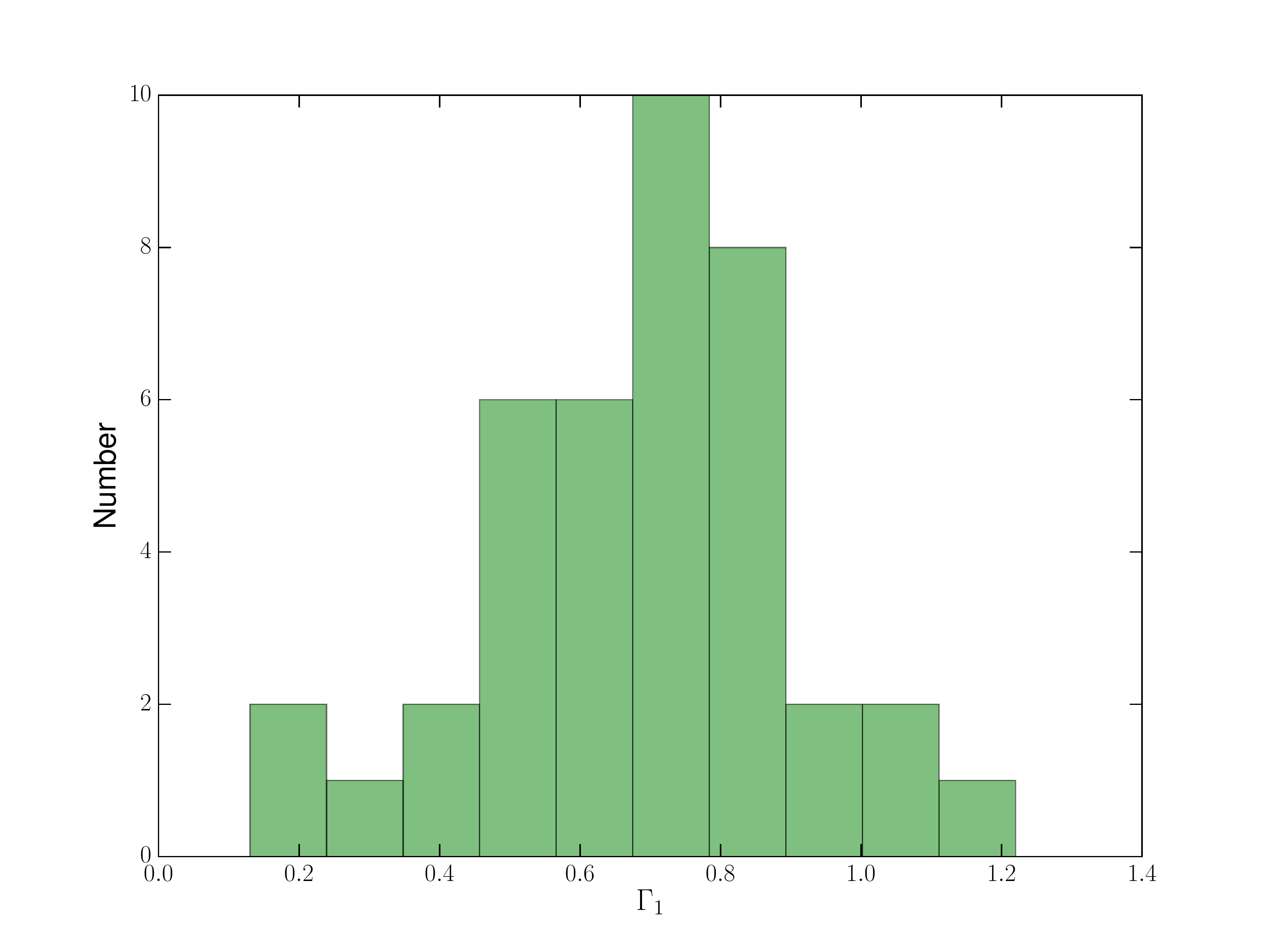}
\caption{(top:) The best-fit results of the simple power law for entire sample of 40 low-mass systems (the entropy is given in keV $\times$ cm$^2$; see text for more details). (bottom:) The histogram of best fit slopes for the low-mass sample.}
\label{fig_hist}
\end{figure}

\begin{figure*}
\centering
\includegraphics[width=0.49\textwidth]{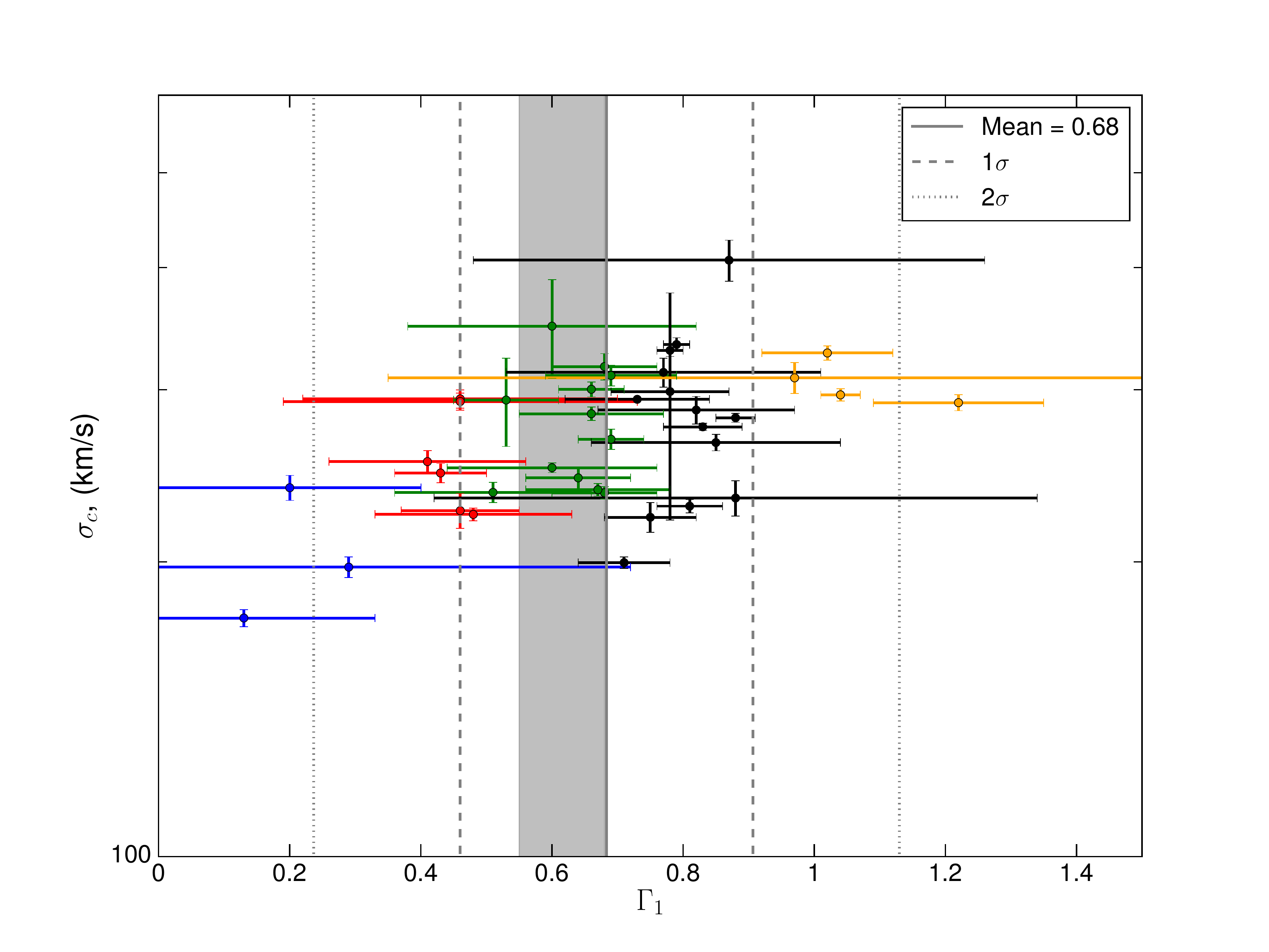}
\includegraphics[width=0.49\textwidth]{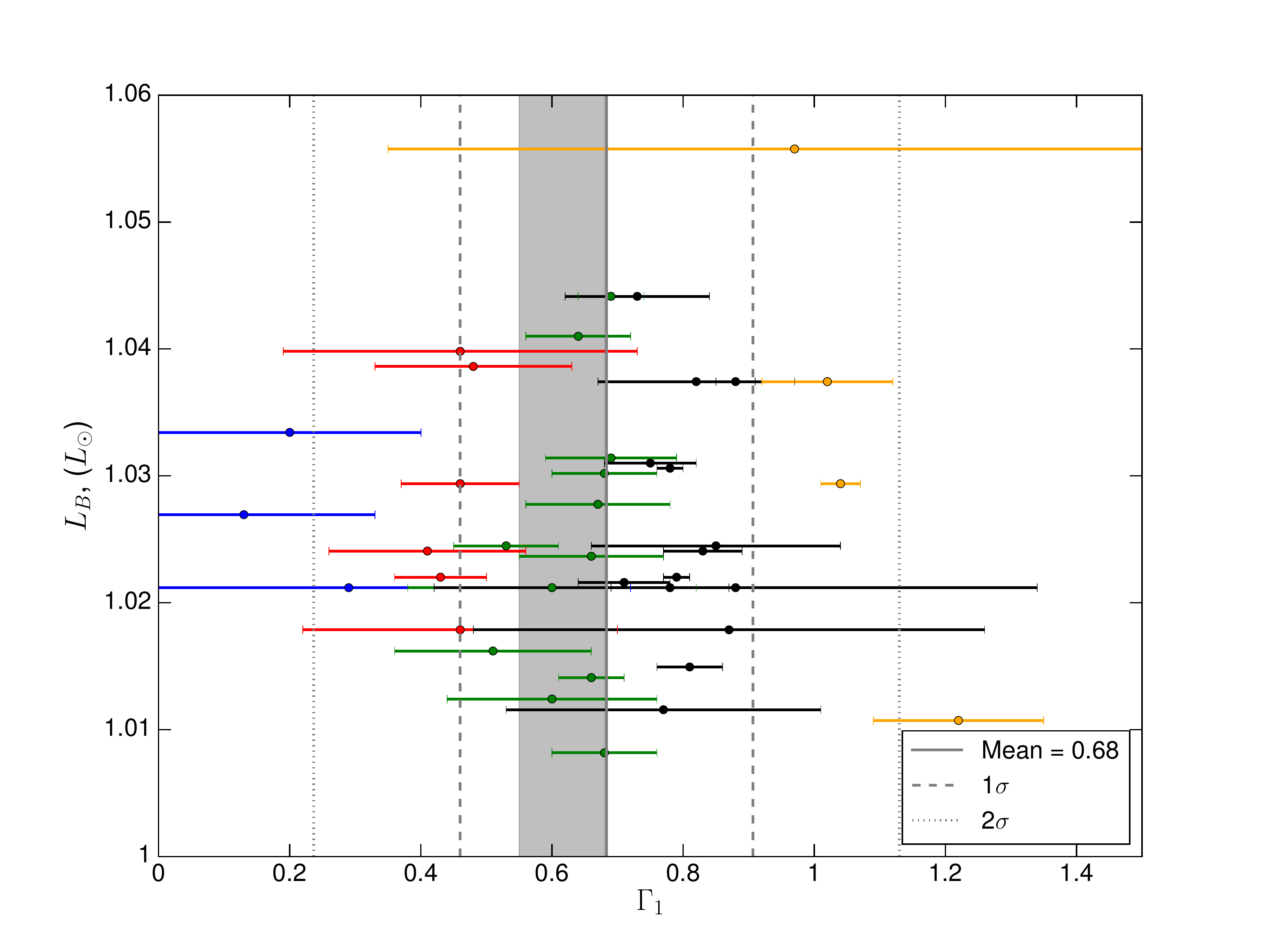}
\includegraphics[width=0.49\textwidth]{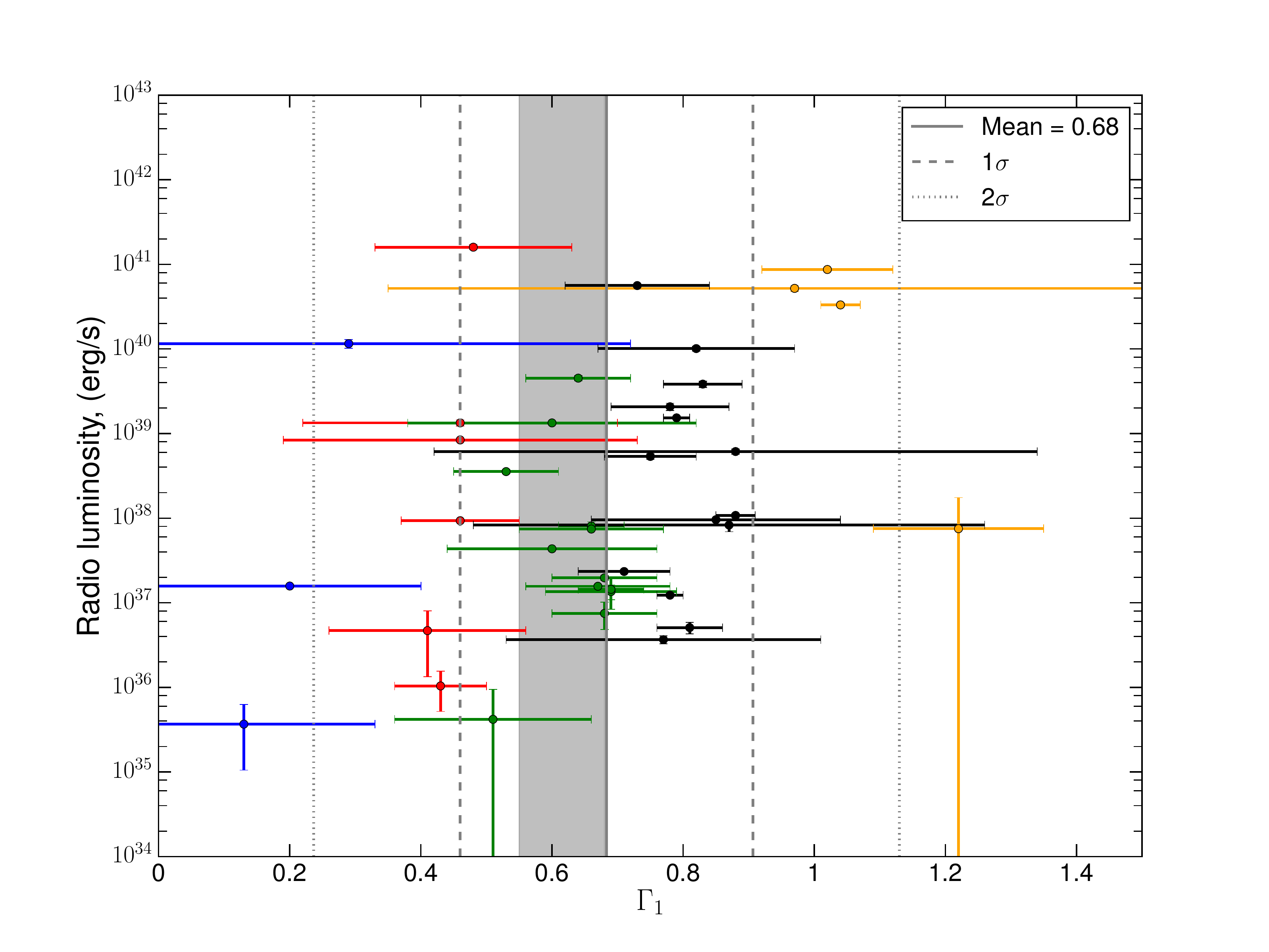}
\includegraphics[width=0.49\textwidth]{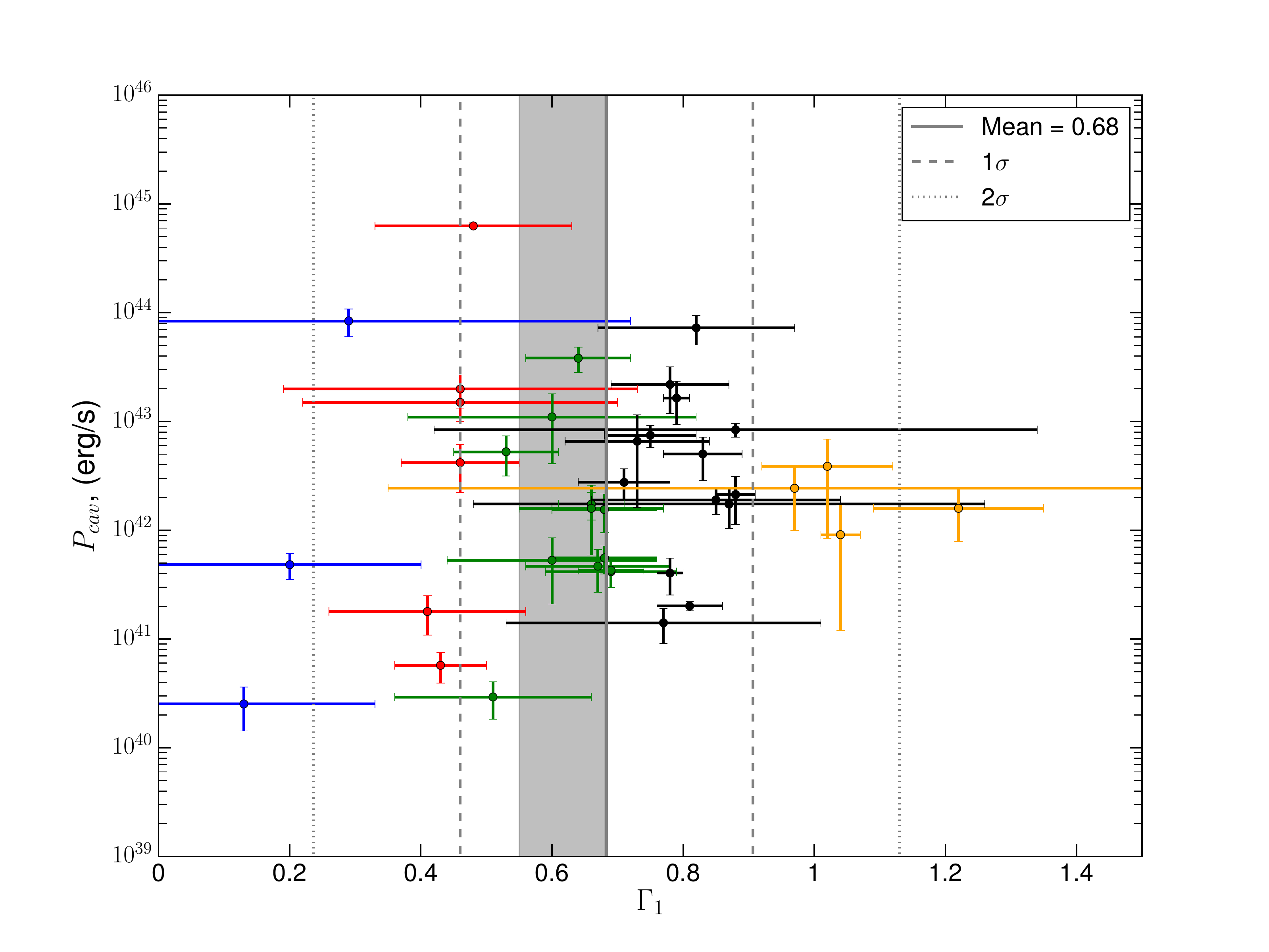}
\caption{The correlations of individual best-fit slopes with central velocity dispersion, optical and radio luminosities as well as cavity power for low-mass systems. The shaded areas correspond to the best-fit result with 1$\sigma$ uncertainty for all 40 systems, while vertical solid lines with dashed and dotted lines correspond the mean and its 1 and 2$\sigma$ confidence levels, respectively. The colors of points are the same as given on previous plot.}
\label{fig_cor}
\end{figure*}

Modeling the entropy profiles individually reveals a relatively broad range of slopes, $\Gamma_1 = 0.2-1.2$. Their distribution is presented in bottom plot of Figure~\ref{fig_hist}. The lowest slopes are obtained for NGC499, NGC3923, NGC4125, NGC4382, and UGC408, while the highest are for NGC533, NGC4104, NGC4261, IC4296, and NGC4782.  
The errors on individual slope measurements shown in Figure~\ref{fig_cor} are  often large compared to the bin size in of the histogram in Figure~\ref{fig_hist}. This indicates that much (but likely not all) of the variance is measurement scatter.  

To evaluate this scatter, we combined these systems into five slope bins: $<$0.3, 0.3--0.5, 0.5--0.7, 0.7-0.9, and $>$0.9, and fit power laws to the means.  In the upper panel of Figure~\ref{fig_hist} we show the results for each separate group. The colors  correspond to the individual and binned data as shown. A comparison of the mean profile slopes to the slopes of the individual systems shows that the outlying bins above and below 2/3 are populated primarily by noisy data.  For example, the mean slope for the objects falling in the bin 0.1--0.3 includes NGC4125, NGC4325, and UGC408 with power-law slopes 0.20$\pm$0.20, 0.15$\pm$0.30, 0.29$\pm$0.43, respectively.  For these the errors on the power law models are large.  This is similarly true in the highest bin $>$0.9.  When averaged, the slope declines to  $0.82 \pm 0.03$. The upshot is that averaging brings the noisy data closer to the mean value of 2/3. The average slope for these five groups is 0.65$\pm$0.06, consistent with our earlier analysis and consistent with \citet{Panagoulia:14} and \citet{Hogan:17a}. 

Despite the very strong evidence for a universal inner profile, a few systems with relatively small errors may depart significantly from the mean. To explore this further we correlated the slopes of individual objects with several physical parameters that may cause the slopes to deviate from the mean.  

We explore trends between entropy slope and several physical properties of these systems including  central velocity dispersion, optical luminosity, radio flux and luminosity, and AGN cavity power in Figure~\ref{fig_cor}. The velocity dispersions were taken from HyperLEDA while the optical luminosities were taken from \citet{Ellis:06, Nagino:09}.  Radio fluxes were taken from NED for 1.4 GHz VLA data and converted to radio luminosity using $L = S \times 4\pi \times D_L^2 \times \nu$.  Here $S$ is the radio flux, $D_L$ is the luminosity distance, and $\nu$ is the frequency.  The radio luminosities were converted to mechanical feedback cavity power using the \citet{Cavagnolo:10} scaling relation.  

Figure~\ref{fig_cor} shows the trends.  No clear correlation emerges between entropy profile slope and radio luminosity, mechanical power, stellar velocity dispersion, or galaxy luminosity.  For reference the mean slope and its $1\sigma$ and $2\sigma$confidence intervals are shown.   The best-fit power law slope 0.62$\pm$0.09 obtained for all low-mass systems, with 1$\sigma$ uncertainty are indicated by the shaded areas in Figure~\ref{fig_cor}.

The ranges of velocity dispersion and optical luminosity are limited and thus have little leverage.  However the range of radio luminosity spans $\sim$ 6 orders of magnitude. This is noteworthy because AGN outbursts can have a dramatic effect on hot atmospheres as they deposit enormous amounts of energy.  Yet even the largest outburst have little impact on the central gas density profiles \citep{Hogan:17a,McNamara:16}.  This is also true for entropy.  Central entropy profile slopes remain close to $R^{2/3}$ regardless of AGN power.  This is inconsistent with many feedback models (cf., \citet{Hogan:17a, Pulido:17}).

%\citet{Werner:14} indicated that low-mass systems with entropy slopes 0.9--1.0 harbour reservoirs of cold gas and extended optical-emission line nebulae. They showed that systems with such cold gas characterized by the entropies which are systematically lower at $R \geq$ 1 kpc than for those systems without cold gas. We compared our
%slopes to the five objects in common with Werner:   0.79$\pm$0.02 for NGC1399, 0.85$\pm$0.19 for NGC1407, 1.04$\pm$0.03 for NGC4261, 0.88$\pm$0.03 for NGC4472, and 0.78$\pm$0.02 for NGC4649.  Our slopes agree with \citet{Werner:14} but with large uncertainties. 

\subsection{Entropy profiles for S0 and S galaxies} 

\begin{figure}
\centering
\includegraphics[width=0.24\textwidth]{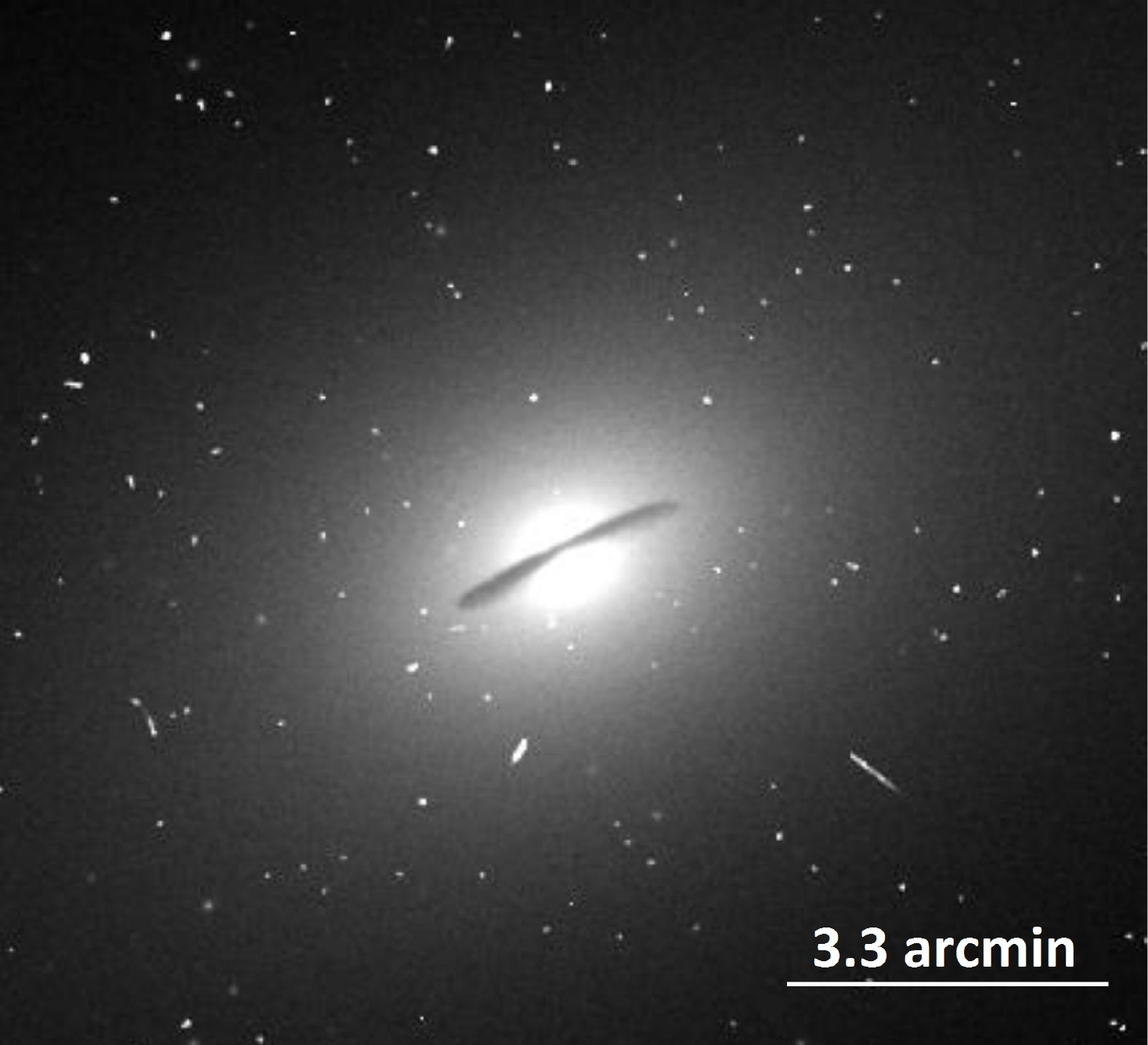}
\includegraphics[width=0.231\textwidth]{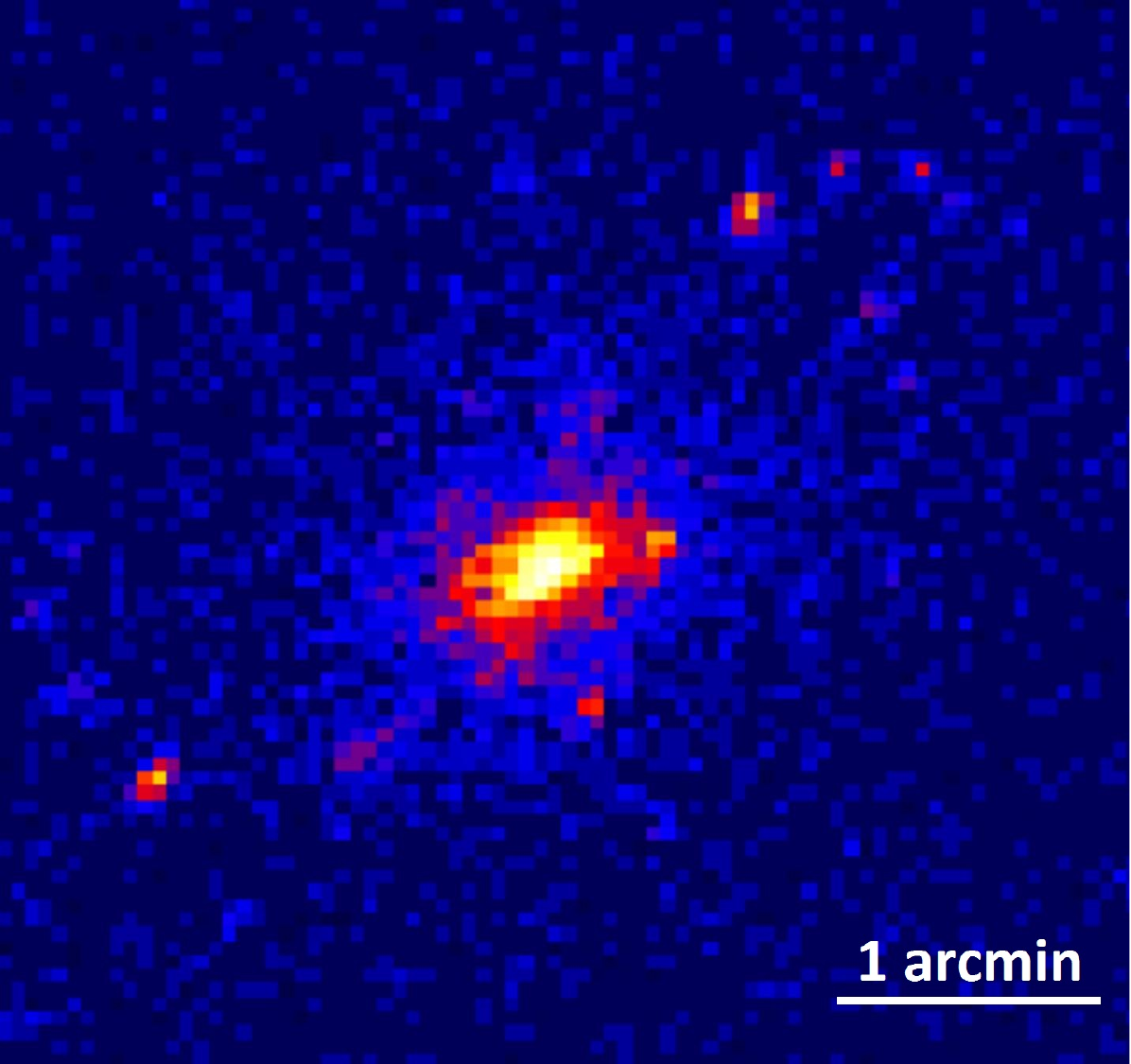}\\
\includegraphics[width=0.24\textwidth]{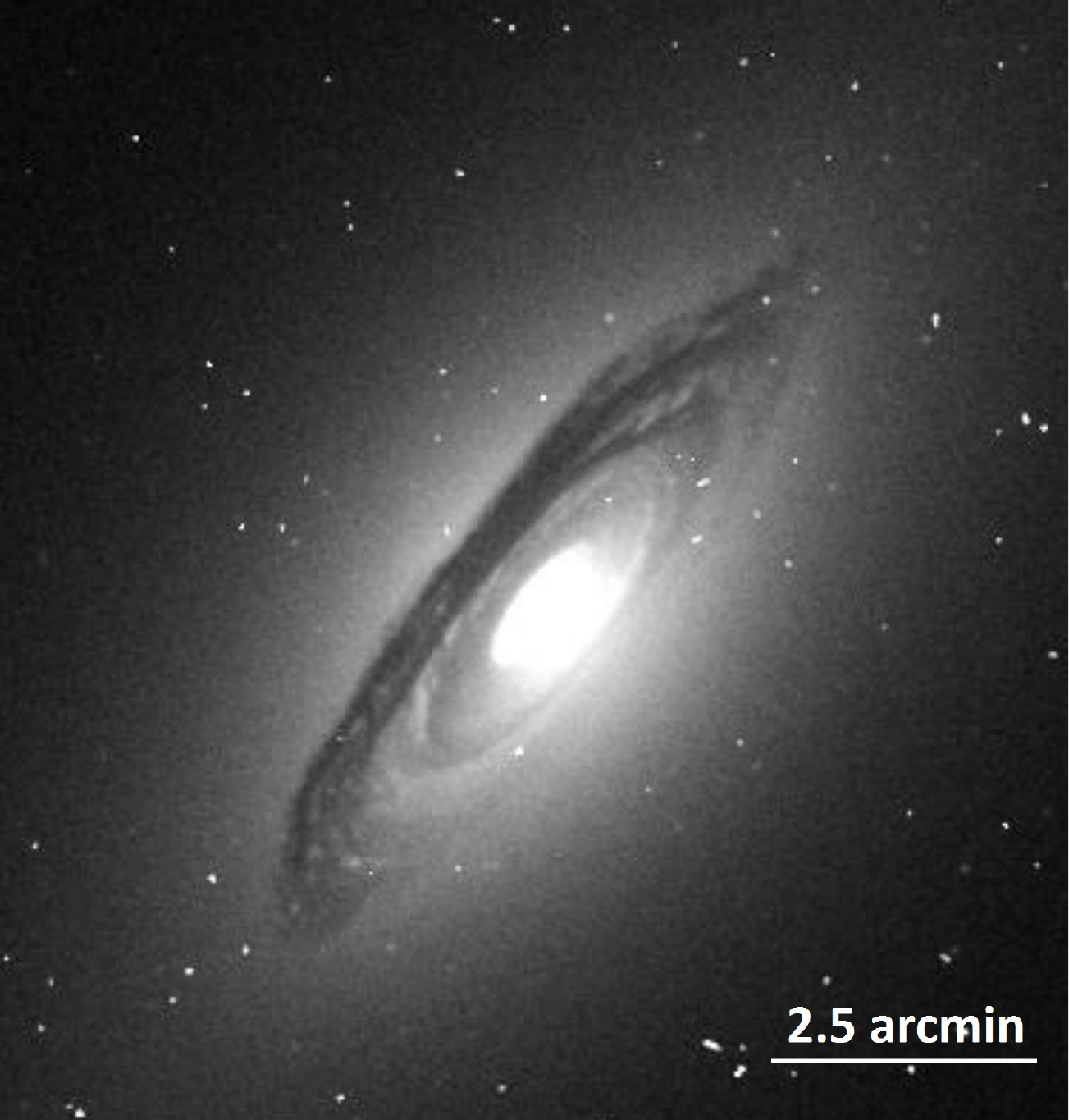}
\includegraphics[width=0.233\textwidth]{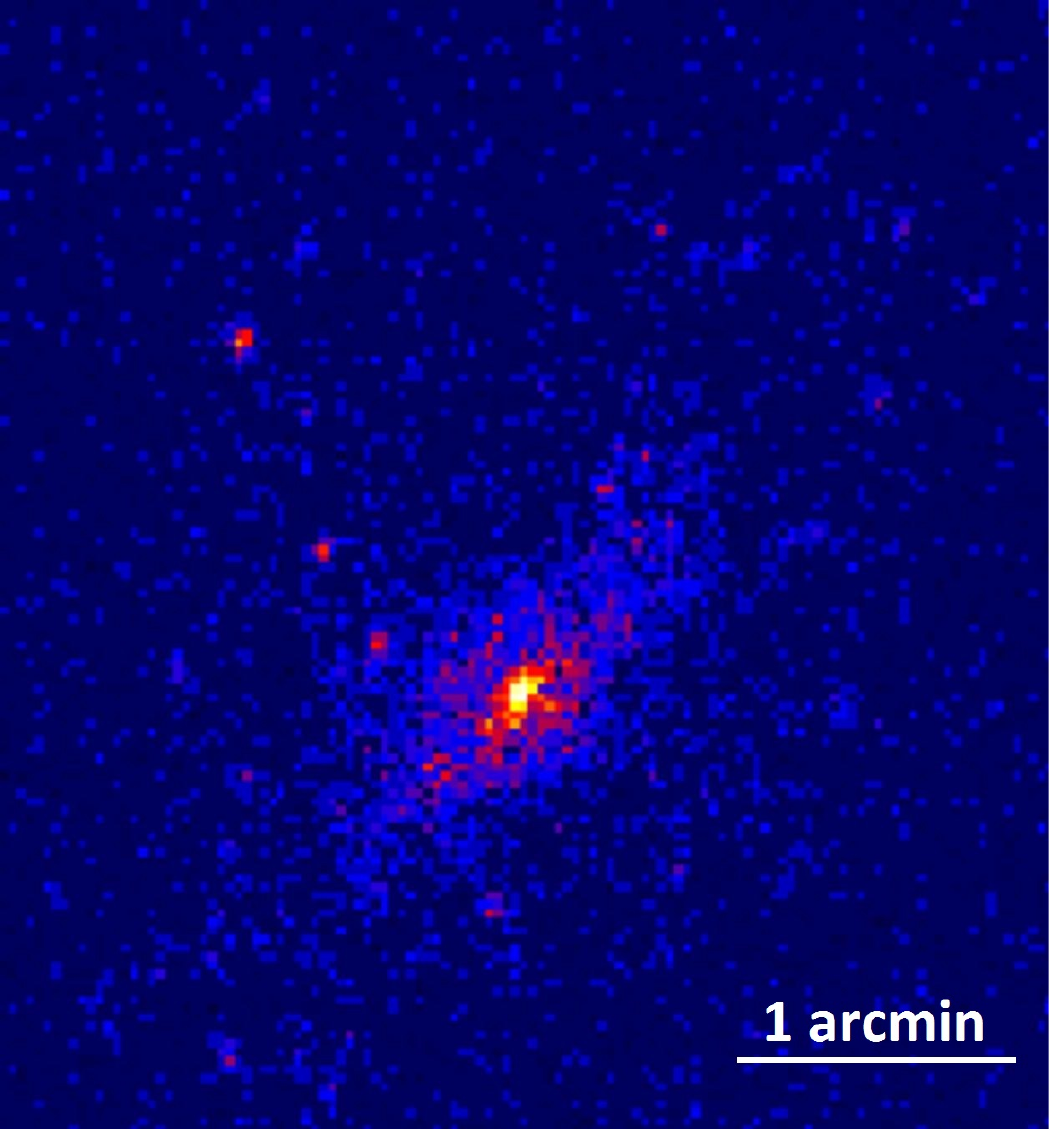}
\caption{5439.0 $\text{\AA}$ Hubble Space Telescope (left) and 0.5--7.0 keV $Chandra$ (right) images of the S0 (NGC1332) and SA0 (NGC6861) galaxies.}
\label{fig_im1}
\end{figure}

The six lenticular and spiral galaxies in our sample have prominent dust disks. X-ray and optical images of two of the most prominent lenticular and spiral galaxies are shown in Figure~\ref{fig_im1}.

\begin{figure}
\centering
\includegraphics[width=0.49\textwidth]{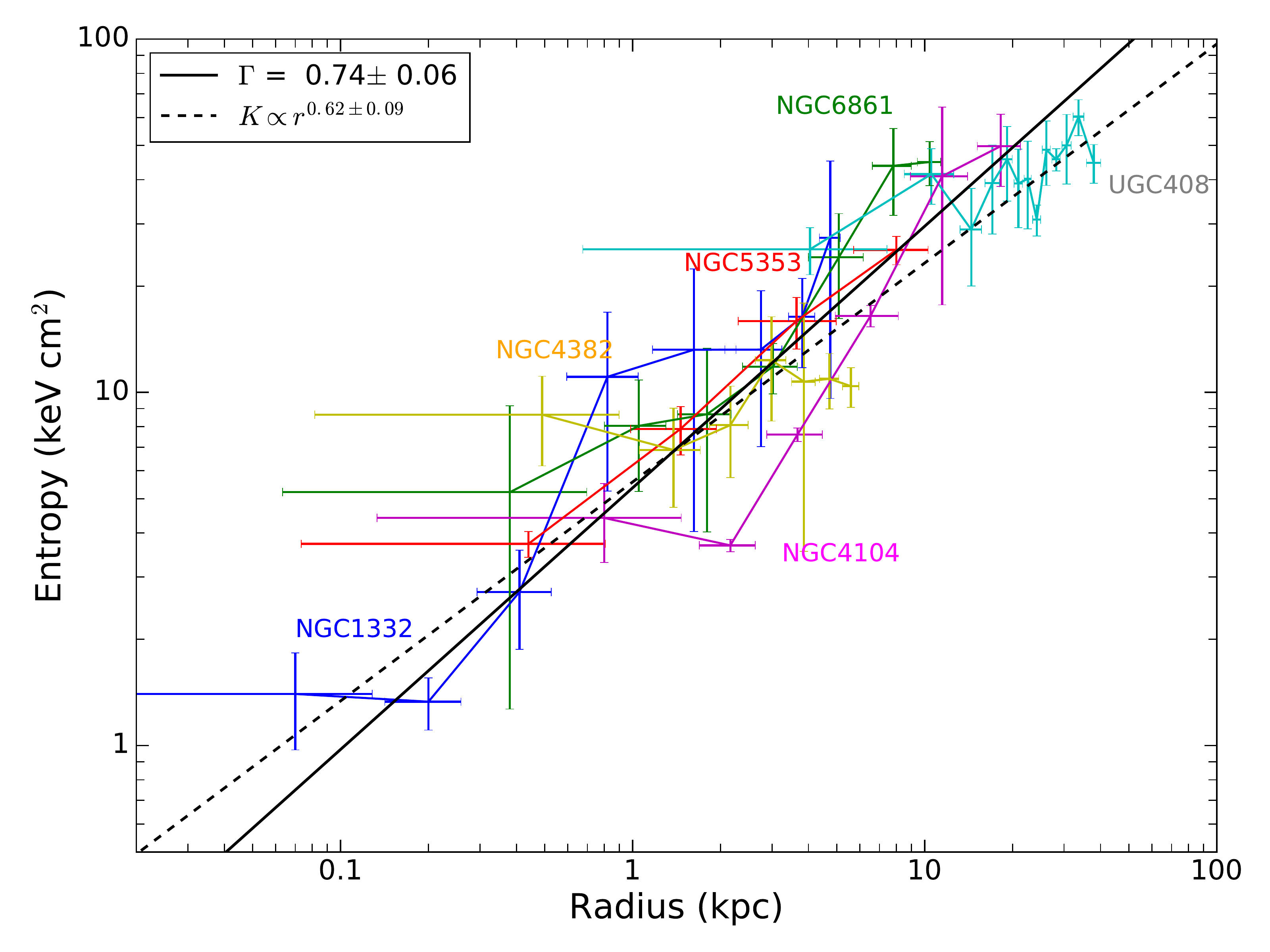}\\
\caption{Deprojected entropy profiles with the best-fit slope (solid line) for the lenticular and spiral galaxies. The best-fit slope for the entire sample (dashed line) is shown for comparison.  The S0 and S galaxies follow the same slope as the ellipticals.}
\label{fig_im2}
\end{figure}

Despite being spiral galaxies, they are bright X-ray sources that permit their entropy profiles to be measured.  Their entropy profiles shown in Figure~\ref{fig_im2} have similar slopes ($\Gamma = 0.74\pm0.06$) to the elliptical and brightest cluster galaxies ($\Gamma = 0.62\pm0.09$). This is significant as it links the thermodynamic properties of their atmospheres across galaxy morphology  and halo mass.  It indicates that the underlying physics that imprints the entropy distributions in elliptical and spiral galaxies is 
similar. 

\section{Discussion}

Consistent with many previous studies, we have shown that the entropy profiles of galaxies and clusters can be described by a broken power law.  The inner profiles for central galaxies of all types, including spirals,  is characterized by $K\propto R^{2/3}$, with only small variations.  At large radii, between $\sim 0.1-1 R_{2500}$, $K\propto R$. The gas entropy in the outer parts of cluster atmospheres imprinted primarily by gravitational collapse is expected to follow $K \propto R^{1.1}$ \citep{Voit:02, Voit:03, Voit05, Cavagnolo:08}.  Our outer profiles are slightly shallower. 

The $K \propto R^{2/3}$  inner slope reflects higher gas entropy than expected from an inward extrapolation of the $K \propto R^{1.1}$ profile.  Higher central gas entropy indicates additional heating of the atmosphere in the vicinity of the central galaxy.  Recent studies have shown that when resolution effects are properly accounted for, the inner entropy profile in clusters can be described as $K \propto R^{2/3}$ \citep{Panagoulia:14, Hogan:17a}.  The $R^{2/3}$ form is significant as we shall see below.  
%We now explore whether the entropy profiles for early-type galaxies in the central regions of halos less massive than clusters follow the same form.

How does this broken power-law form arise? When hot atmospheres are heated by radio jets launched by central black holes or by supernova explosions, they do not necessarily respond with a large temperature rise. Instead, heating raise the entropy of the gas, causing it to expand and lift outward.  Since gravity is weaker at larger radii, the weight of the gas is reduced causing the pressure to decrease.  As a result, most of the heat energy is converted into gravitational potential energy, rather than thermal energy. Conversely, as the atmosphere cools and its entropy decreases, the gas contracts and moves inward, again with little change in the gas temperature.  Under the right conditions, some thermally unstable gas can cool faster than the rest, condensing into molecular clouds that form stars and feed the nuclear black hole. The entropy parameter encodes information about the heating and cooling history of hot atmospheres.  

The break to a shallower inner slope seen in Figure~\ref{fig_1} may mark the boundary, within which the atmosphere is strongly heated by the radio jets. Remarkably, early-type galaxies including spirals and central galaxies of massive clusters follow the same $R^{2/3}$ form.   This indicates that the
$R^{2/3}$ form is linked to the central galaxy.

Assuming the central galaxy is an isothermal sphere, i.e., $M = 2\sigma^2 R / G$, the free-fall time is $t_{\rm ff} = R / \sigma$.  Cooling time scales as $t_c \propto K^{3/2} / (\Lambda \sqrt{kT})$ so, for $K \sim R^{2/3}$, the ratio of cooling time to free-fall time scales as $t_c/t_{\rm ff} \propto \sigma / (\Lambda \sqrt{kT})$. Here, $\sigma$ is the stellar velocity dispersion, $R$ is the radius, $G$ is the gravitational constant, and  $\Lambda$ is the cooling function \citep{Hogan:17a}. This expression has no radial dependence, which is noteworthy.  It implies that $t_c/t_{\rm ff}$ becomes constant where gas becomes thermally unstable. 

The ratio $t_{\rm c}/t_{\rm ff}$ is understood to be related to the condition leading to thermally unstable atmospheric cooling.  
%This occurs inexorably when $t_{\rm c}/t_{\rm ff}\lesssim 1$ \citep{Nulsen:86, Pizzolato:05}.  
%Some studies have suggested that condensation ensues when $t_{\rm c}/t_{\rm ff}\lesssim 10$ \citep{Gaspari:12, Voit:15}. 
The cooling time is defined as the time it takes for atmospheric gas to radiate away its thermal energy. It is expressed here by
\begin{equation}
t_{\rm c} = \frac{3p}{2n_en_H\Lambda(Z, T)} \approx \frac{3pV}{2L_X},
\end{equation}
where $p = 2 n_e k_B T $ is the gas pressure, $\Lambda(Z, T)$ is the cooling function, depends on metallicity and temperature, and $L_X$ is the X-ray luminosity. The free-fall time was given by
\begin{equation}
t_{\rm ff}(r) \simeq \sqrt{2r/g},
\end{equation}
where $g = (G M)/r^2$ is the local gravitational acceleration and the total mass, $M$, was taken from our mass profiles \citep{Babyk:17prof}. {
%A detailed discussion of $t_{\rm c}/t_{\rm ff}$ threshold is presented in \citet{McNamara:16}, \citet{Hogan:17a}, and \citet{Pulido:17}. 

Hot atmospheres are expected to become thermally unstable to linear density perturbations when the ratio of \tctff\ falls below unity \citep{Nulsen:86, Pizzolato:05, McCourt:12}. This criterion is never achieved in static hot atmospheres.  Nevertheless, molecular gas and star formation are observed in central cluster galaxies indicating that thermally unstable cooling is occurring.  Recent studies have suggested that this instability criterion may rise well above unity, so that thermally unstable cooling can ensue from linear perturbations when \tctff\ falls below 10 \citep{Sharma:12, McCourt:12, Gaspari:13, Li:15, Voit:15a}.  \citet{McCourt:12} argued that this condition was met in systems whose central galaxies contained significant levels of cold gas.   

However, more recent analyses of cluster central galaxies paying close attention to mass profile measurements and resolution effects  have revealed no evidence that \tctff\ falls significantly below 10 in any system, including those that are  thermally unstable \citep{McNamara:16, Hogan:17a, Pulido:17}.  Instead,  these studies found that \tctff\ lies between about 10 and 30 in systems with star formation and molecular clouds.  In addition, lower values of \tctff\  do not correlate with higher star formation rates or molecular gas masses. Consistent with \citet{Voit:15}, they found a floor at \tctff\ $\sim$ 10 rather than a threshold. While this floor may well be physically significant, the range of \tctff\ values can be explained as an observational selection effect \citep{Hogan:17}. Furthermore, newer simulations have shown that the physical bases for the \tctff\ criterion is invalid, and that thermally unstable cooling may occur over a much larger parameter space \citep{Sharma:16}. While it is clear that the cooling time of the hot atmosphere is correlated with the presences or absence of thermally unstable cooling, the ratio \tctff  in clusters does not.  Here we perform a similar analysis on the atmospheres of giant elliptical galaxies.

} 

\begin{figure}
\includegraphics[width=0.49\textwidth]{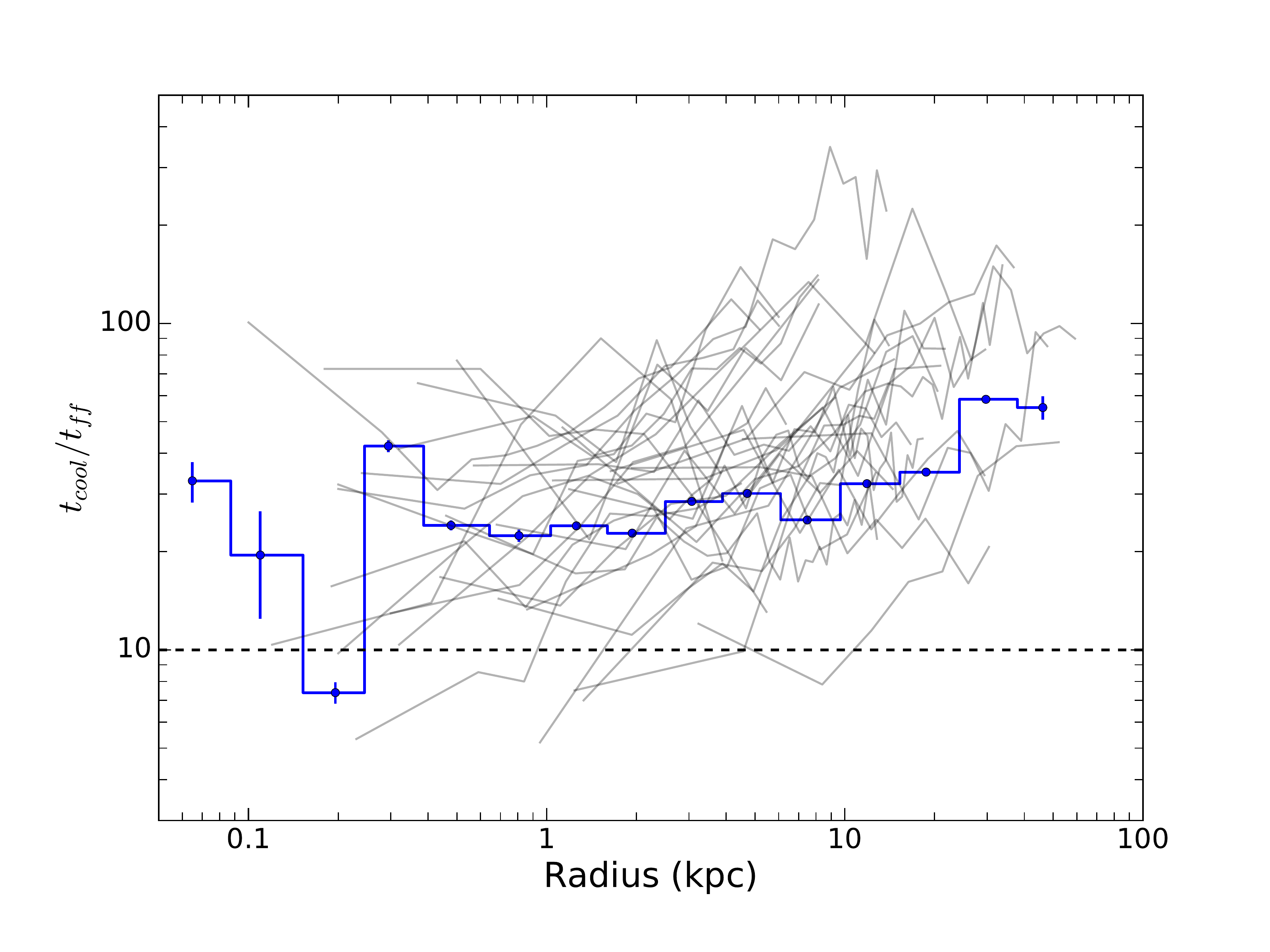}
\caption{Deprojected $t_c/t_{\rm ff}$ profiles (gray shaded lines) for entire sample of low-mass systems along with the average $t_c/t_{\rm ff}$ profile (blue line). The error bars of $t_c/t_{\rm ff}$ profiles have been omitted for clarity.}
\label{fig_tctfprof}
\end{figure}

The ratio $t_{\rm c}/t_{\rm ff}$ for objects in our sample is shown in Figure~\ref{fig_tctfprof}. The $t_{\rm c}/t_{\rm ff}$ profiles were binned in the same way as the entropy profiles, and an average profile was computed. We see that average profile is constant over all radii, with the exception of few outliers seen in the first couple bins. This is likely resolution effect. Nevertheless, the minimum $t_{\rm c}/t_{\rm ff}$, including the average profile, never falls significantly below 10 in the targets, but instead lies between 10 and nearly 100. The average profile lies close to 30, which is in agreement with $t_{\rm c}/t_{\rm ff}$ profiles obtained for galaxy clusters \citep{Hogan:17a, Pulido:17}. Analyses that properly account for the gravity of the central galaxy and for resolution biases, including the data analyzed in \citet{Werner:12, Voit:15, Hogan:17a, Pulido:17}, have found that $t_c/t_{\rm ff}$ always exceeds 10.  This is true in atmospheres that are demonstrably cooling into molecular clouds.  Therefore, $t_c/t_{\rm ff}\lesssim 10$ is not a threshold for thermal instability \citep{Hogan:17a, Pulido:17, Babyk:17prof}. It instead indicates the degree to which the bulk of the atmosphere is thermodynamically stable.  

The closer this ratio is to unity, the more susceptible the atmosphere becomes to thermally unstable perturbations.  For example, perturbations introduced as gas is lifted by radio bubbles are linked to thermally unstable cooling \citep{Pizzolato:05, McCourt:11, McNamara:16}.  Atmospheric gas lifted outward by a galaxy collision or turbulence may also trigger thermally unstable cooling.  For the inner region where the entropy varies as $R^{2/3}$,  $t_c/t_{\rm ff}$ is approximately constant.  This is the region most susceptible to perturbations that promote thermally unstable cooling.  

The entropy slope within $\sim 0.1R_{2500}$  indicates that the cooling time of the atmosphere, which rises with radius, is in close balance with the heating timescale, i.e., $t_c \sim t_H$.   As thermally unstable gas cools and leaves the hot atmosphere, the mean entropy of the remaining hot gas rises.  The cooling gas then fuels the active nucleus which raises the entropy leading to the inner floor in $t_c/t_{\rm ff}$.  At larger radii where the entropy reaches $K \sim R^{1.1}$, the cooling time exceeds $10^9$ yr.  The slightly shallower slope we find indicates that
heating by AGN may be important at altitudes approaching $R_{2500}$.    

The break radius of $0.1R_{2500}$ corresponds to linear scales of $20-40$ kpc in central cluster galaxies, which is indeed the region where thermally unstable cooling leads to nebular line emission \citep{McDonald:11}, molecular gas \citep{Edge:02}, and star formation proceeding at tens of solar masses per year \citep{ODea:08}.  However, this radius is smaller,  $\sim 15$ kpc, in early-type galaxies. Unlike central cluster galaxies, early-type galaxies contain much lower levels of molecular gas. They rarely form stars at significant rates \citep{Werner:14} despite having similar entropy profiles.

\subsection{Why are Elliptical Galaxies Dormant?}

Why most early-type galaxies lie dormant, despite short central cooling times, while some central cluster galaxies are burgeoning is puzzling.  The reasons are twofold.  First, the hot atmospheres of early-type galaxies/groups contain less mass than those in clusters.  Therefore, their fuel reservoirs are smaller.  Second, the active nuclei in lower mass, early-type galaxies supply more  energy per gas particle than the active nuclei of central cluster galaxies. Central cluster galaxies with cooling atmospheres have typical jet powers and gas masses within 0.1$R_{2500}$ of $P_{\rm jet} \approx 10^{43}$ erg/s and $M_g \approx$ 10$^{12} M_{\odot}$, respectively. The energy absorbed per gas particle, $\epsilon = \eta E_{\rm tot} \mu m_p / M_{g}$ \citep{Ma:11}, is then $\sim0.1$ keV/particle.  Here $\mu=0.63$ is the mean molecular weight of the gas.  $\eta \simeq 0.1$ is an efficiency factor that accounts for the fraction of the jet's enthalpy that heats the atmosphere and the fraction of the bubble enthalpy deposited in the inner region \citep{Weinberger:17}. In many systems, this level of heating cannot quite keep up with the rate of cooling, leading to significant star formation.  
 
On the other hand, for early-type galaxies/groups with average jet powers of $P_{\rm jet} \approx$ 10$^{41-42}$ erg/s, and atmospheric gas masses, $M_{g} \approx$ 10$^{9} M_{\odot}$  within 0.1$R_{2500}$, we find an average heating level of $\sim$ 1-10 keV/particle. We have further assumed all systems have been active at the same level for 10 Gyr, and we have ignored radiative cooling. 
%{\bf Moreover, using Euler's method we numerically integrate the broken power law model within first and last observable entropies. During our integration we assume the best-fit parameters of broken power law model obtained above. We also do the same integration for a simple power law model assuming the best-fit slope, $\Gamma_2$, and normalization obtained for the broken power law model, to define excess of entropy, i.e. additional heating. Thus, we divide the power law result from the broken power law and measure the entropy excess. To convert the obtained entropy excess from keV per cm$^2$ to keV per particle we use the values of column density, $N_H$ (particle per cm$^2$), for each sampled system individually given in Table~\ref{tab1}. The distribution of obtained heating per particle calculations is shown as a histogram in Figure~\ref{fig_hpp}. As clearly seen from histogram, the excess of heating is distributed from few to several keV per gas particle. The measured excess is at least an order of magnitude higher than those measured for clusters. This result confirms recent and our calculations of heating per gas particle. 
%{\bf We also find an excess of entropy at 1 kpc as 6 keV cm$^2$ that agrees with our calculations above as well.} 

While crude, these calculations indicate that the level of heating per gas particle by active nuclei in early-type galaxies generally exceeds that in groups and clusters.  Apparently active nuclei in early-type galaxies are better able to prevent significant levels of star formation while allowing enough cooling near the nucleus to maintain the energetic feedback loop.

The similarity in entropy profile shape across such an enormous range of jet power, halo mass, and atmospheric gas mass is significant.  It indicates that the AGN feedback mechanism is extraordinarily responsive and gentle.  At the same time, the atmosphere is able to maintain a rough balance between heating and cooling throughout the entire $K\propto R^{2/3}$ cooling region. This represents a deep challenge to hydrodynamic simulations, which generally show dramatic time variations in the temperature, density, and entropy profiles in response to radio jet interactions \citep{Gaspari:12, Sijacki:11}. The active nucleus must respond promptly to heating and cooling over the entire $0.1R_{2500}$ region without inducing large entropy excursions in response to the jet \citep{Gaspari:12, Sijacki:11, Weinberger:17}. This is a general result that extends throughout the elliptical galaxy mass range including spiral galaxies.  

% \begin{figure}[ht!]
% \centering
% \includegraphics[width=0.48\textwidth,natwidth=610,natheight=642]{heat.pdf}
% \caption{The histogram of heating per gas particle values for entire sample of low-mass systems.}\label{fig_hpp}
% \end{figure}

%{\bf Extra text} The best-fit result of this fitting confirms our recent estimates for all clusters. Our results for non-cool-core and cool-core clusters are consistent with previous studies \citep{David:96, Ponman:99, David:01, Ponman:03, Panagoulia:14, Hogan:17a, Pulido:17}. Previous $Chandra$ and $XMM-Newton$ studies reveal the flattening of inner entropy profiles in clusters. \citet{Donahue:05, Donahue:06, Cavagnolo:09, Voit:16} applied functional fit of $K(r) = K_0 + K_{100}(r/100{\rm kpc})^{\alpha}$. This form is good representation of clusters with high central entropy only \citep{Cavagnolo:09}. In contrast, this form represents cool-core clusters poorly. \citet{Panagoulia:14} suggest that the broken power law fit where the inner entropy follows $K \propto r^{0.67}$ is a good representation for cool-core clusters. The value of inner slope, 0.67, agrees well with our measurements. Below we provide the physical implication and detailed analysis of this slope.

\section{Summary}
We have presented new entropy profiles of 40 elliptical galaxies, spiral galaxies, and faint groups of galaxies, and we compared them to 110 entropy profiles of cool-core and non-cool-core galaxy clusters observed by $Chandra$ X-ray Observatory. The entropy profiles cover a wide range of temperature, total mass, gas mass, radio, and central galaxy optical luminosity, and jet power. Our results are summarized as follows:
\begin{itemize}
\item The entropy profiles within $\sim$ 0.1$R_{2500}$ of elliptical, lenticular, early spiral, and brightest cluster galaxies, follow approximately as $K \propto R^{2/3}$.
\item Beyond 0.1$R_{2500}$  entropy profiles are slightly shallower than $K \propto R^{1.1}$, indicating that heating, likely by AGN feedback, extends well beyond the central galaxy. 
\item {This sample includes 22 non-cool core clusters. Four are centered on a bright galaxy; 18 are not. The central entropy values of non-cool core clusters with and without BCGs lie above $\sim$ 50 keV cm$^2$. Their entropy profiles, though not well defined due to poor photon statistics, are shallower than the  $R^{2/3}$ form seen in cool-core clusters that contain central galaxies. The outer entropy profiles of non-cool core clusters follow the $R^{1.1}$ profile beyond $\sim$ 0.1$R_{2500}$, similarly to cool core clusters.  It is not clear how the $R^{2/3}$ form arises. Does it depend on the central galaxy alone or the existence of a central galaxy residing in a cooling atmosphere stabilized by feedback? The data thus far suggest the latter. Additional study is needed to clarify the issue. 
%Our sample includes 22 non-cool core clusters that host and do not a central galaxy. The central entropies of non-cool core clusters with/without BCGs lie above $\sim$ 50 keV cm$^2$ and do not follow the $R^{2/3}$ form seen in cool-core clusters that contain central galaxies. However, their central entropy profiles are similar and demonstrate flattening which is poorly defined because of poor statistics. The outer entropy profiles of non-cool core clusters follow the $R^{1.1}$ profile beyond $\sim$0.1$R_{2500}$ similarly to cool core clusters. 
%Therefore the primary difference between the cool core and non-cool core clusters here is the presence or absence of a central galaxy. The $K \propto R^{2/3}$ form appears to be linked to the central galaxy.
}
\item The $K \propto R^{2/3}$ entropy profile shape is intimately related to the central galaxy itself and is consistent with thermally unstable cooling balanced by heating where the inner cooling and free-fall timescales approach a constant ratio.
\item Hot atmospheres of early-type galaxies are heated at a higher rate per gas particle than
central cluster galaxies. The extra heating may explain at least in part why early-type galaxies are largely dormant.
\end{itemize}

\acknowledgements
BRM acknowledges funding from the Natural Science and Engineering Research Council of Canada and from the Canadian Space Agency. Support for this work was provided in part by the National Aeronautics and Space Administration through Chandra Award Number G07-18104X (Cen A) issued by the Chandra X-ray Observatory Center, which is operated by the Smithsonian Astrophysical Observatory for and on behalf of the National Aeronautics Space Administration under contract NAS8-03060. The scientific results reported in this article are based on observations made by the Chandra X-ray Observatory and has made use of software provided by the Chandra X-ray Center (CXC) in the application packages CIAO, ChIPS, and Sherpa. This research has also made use of the NASA/IPAC Extragalactic Database (NED).\\

\bibliographystyle{apj}
\bibliography{paper}

\end{document}